\pdfoutput=1
\documentclass[twocolumn]{aastex631}
\usepackage{graphicx}
\usepackage{float}
\usepackage{placeins}
\usepackage{xcolor}
\usepackage{natbib}
\usepackage{amsmath}
\usepackage{hyperref}

\pagecolor{white}
\begin{document}

\title{The AURORA Survey: Constraining Chemical Enrichment Pathways at Cosmic Noon with Argon Abundances}
\author[0009-0005-0100-7612]{Jack Foley}
\affiliation{Department of Physics and Astronomy, University of California, Los Angeles, 430 Portola Plaza, Los Angeles, CA 90095, USA}
\author[0000-0003-3509-4855]{Alice Shapley}
\affiliation{Department of Physics and Astronomy, University of California, Los Angeles, 430 Portola Plaza, Los Angeles, CA 90095, USA}
\author[0000-0003-4792-9119]{Ryan Sanders}
\affiliation{Department of Physics and Astronomy, University of Kentucky 505 Rose Street, Lexington, KY 40506, USA}
\author[0000-0001-9687-4973]{Naveen A. Reddy}\affiliation{Department of Physics \& Astronomy, University of California, Riverside, 900 University Avenue, Riverside, CA 92521, USA}
\author[0000-0001-8426-1141]{Michael W. Topping}\affiliation{Steward Observatory, University of Arizona, 933 N Cherry Avenue, Tucson, AZ 85721, USA}
\author[0000-0002-0827-9769]{Thomas M. Stanton}
\affiliation{Institute for Astronomy, University of Edinburgh, Royal Observatory, Edinburgh EH9 3HJ, UK}
\author[0000-0002-5139-4359]{Max Pettini}\affiliation{Institute of Astronomy, Madingley Road, Cambridge CB3 OHA, UK}
\author[0000-0002-3736-476X]{Fergus Cullen}\affiliation{Institute for Astronomy, University of Edinburgh, Royal Observatory, Edinburgh, EH9 3HJ, UK}
\author[0000-0001-7782-7071]{Richard S. Ellis}\affiliation{Department of Physics \& Astronomy, University College London. Gower St., London WC1E 6BT, UK}
\author[0000-0003-4264-3381]{N. M. F\"orster Schreiber}\affiliation{Max-Planck-Institut f\"ur extraterrestrische Physik (MPE), Giessenbachstr.1, D-85748 Garching, Germany}
\author[0000-0001-5860-3419]{Tucker Jones}\affiliation{Department of Physics and Astronomy, University of California Davis, 1 Shields Avenue, Davis, CA 95616, USA}
\author[0000-0003-4464-4505]{Anthony J. Pahl}
\affiliation{The Observatories of the Carnegie Institution for Science, 813 Santa Barbara Street, Pasadena, CA 91101, USA}
\author[0000-0003-1249-6392]{Leonardo Clarke}
\affiliation{Department of Physics and Astronomy, University of California, Los Angeles, 430 Portola Plaza, Los Angeles, CA 90095, USA}
\author[0000-0001-9489-3791]{Natalie Lam}
\affiliation{Department of Physics and Astronomy, University of California, Los Angeles, 430 Portola Plaza, Los Angeles, CA 90095, USA}

\begin{abstract}

We present argon abundances from a sample of 46 star-forming galaxies at $z=2-3.5$ from the Assembly of Ultradeep Rest-Optical Observations Revealing Astrophysics (AURORA) program. Although argon is an $\alpha-$element produced by Core Collapse Supernovae (CCSNe),  the latest supernova yield models suggest additional argon production and enrichment by Type Ia supernovae (SNe Ia), unlike other $\alpha-$elements such as oxygen. To study the relationship between argon and oxygen abundances, we construct two median-stacked composite spectra for separate $z=2.0-2.6$ and $z=2.8-3.5$ redshift bins, presenting more representative measurements than previous samples that require individual detection of faint lines. Abundance ratios were determined using an empirical calibration based on the strength of the [ArIII]$\lambda7137$ emission line relative to the [OIII]$\lambda5008$ emission line. With this calibration, we estimate argon abundances (Ar/O) of $0.42^{+0.12}_{-0.10}\mbox{(Ar/O)}_{\odot}$ for the $\langle z \rangle = 2.26$ bin and $0.42^{+0.12}_{-0.11}\mbox{(Ar/O)}_{\odot}$ for the $\langle z \rangle = 3.15$ bin, suggesting minimal SNe Ia and dominant CCSNe enrichment in this sample. Comparison of our abundance measurements of $z\sim 2-3$ AURORA galaxies with chemical evolution modeling of Milky Way stars shows consistency with the Milky Way Bulge component, suggesting a rapid star-formation timescale. However, even larger samples of actively star-forming galaxies with available argon abundances, as well as comparisons between argon abundance and other critical galaxy properties (e.g., specific star formation rate, sSFR) and models (e.g., one tuned specifically to this redshift range) are needed to draw stronger conclusions on the role of argon in galactic chemical enrichment at Cosmic Noon.

\end{abstract}

\section{Introduction}
The diverse metal profile found in the Milky Way (MW) today is the result of many successive generations of stars that fuse hydrogen and helium into heavier metals, and then eject these metals back into the interstellar medium (ISM) upon their death for incorporation into future stars \citep{Kobayashi2020}. Metallicity in the ISM  drives many crucial processes, such as the hardness of stellar emission spectra \citep{Steidel2016, Shapley2019, Sanders2020a, Topping2020, Cullen2021, Clarke2023}, the density of dust grains \citep{Pettini2002, Izotov2006, Asano2013, Jones2018}, and cooling of stellar material \citep{Wang2014}. These conditions in turn determine the characteristics of star-forming regions and the stellar populations they produce \citep{Sanders2024}. Given the many roles of metals in stars and galaxies, it is of utmost importance that their sources are robustly studied and well understood.

By measuring the abundance patterns of metals in the ISM of galaxies, we may place constraints on both galaxy star-formation histories and assembly in the early Universe. Spectroscopic analyses of MW stars suggest that the relative timescales between different supernovae influence the proportions of $\alpha$-elements (e.g., oxygen, sulfur, and argon), with respect to Fe-Peak elements (e.g., iron and nickel) \citep{Wallerstein1962, Tinsley1979, Matteuci1986}. The analysis of abundance patterns in distant($z>1$) star-forming galaxies reveals similar trends over a longer period of cosmic history \citep[e.g.,][]{Kobayashi2020, Topping2020, Stanton2025}. Both Galactic Chemical Evolution (GCE) modeling \citep[e.g.,][]{Kobayashi2000, Kobayashi2020, Nomoto2013, Monty2025} and observations of $z\geq 2$ galaxies  \citep[e.g.,][]{Pettini2002, Steidel2016,Topping2020, Cullen2021, Stanton2024, Welch2025, Curti2025} reveal a deficiency of iron with respect to alpha elements (i.e., a larger $\alpha/\mathrm{Fe}$) at earlier times when compared to solar abundance patterns at present. This deficiency is generally interpreted as being due to the delayed production of iron by Type Ia Supernovae (SNe Ia), as these events have longer timescales than both Core Collapse Supernovae (CCSNe) \citep{Tinsley1979, Matteuci1986, Maoz2012}, and the typical stellar population ages in distant star-forming galaxies \citep[but see][]{Mannucci2006}. This abundance pattern results in harder ionizing spectra for massive stars at a fixed nebular oxygen abundance when compared with star-forming galaxies in the local universe at $\mathrm{z\sim0}$ \citep[e.g.][]{Steidel2016,Topping2020,Runco2021}.

From these timescale considerations, one would expect that actively star-forming distant galaxies have higher proportions of $\alpha$-elements with respect to iron than the Sun. $\alpha$-enhancement has been characterized in distant star-forming galaxies \citep{Shapley2019, Cullen2021, Stanton2024, Shapley2025}, quiescent galaxies \citep{Thomas2005, Conroy2012, Kriek2019, Beverage2023, Jafariyazani2025, Beverage2025}, and damped Lyman-$\alpha$ absorbers at $ 0.6 < z < 3.4$ \citep{Velichko2024}. Given its prevalence in many astrophysical settings, as well as its influence on the development of stellar populations, $\alpha$-enhancement and the $\alpha/$Fe ratio must be well measured and understood for a large population of stars and galaxies of various ages, masses, and cosmic distances. 

However, measurement of direct iron abundances is often difficult. For star-forming galaxies, most iron is depleted into dust grains, making estimates of total iron abundance from gas-phase emission lines inaccurate \citep{Osterbrock1992, Izotov2006, MendezDelgado2024, Shapley2025}. Furthermore, the weak strength of most iron lines makes them difficult to detect in noisy spectra \citep{Rodriguez2005}, requiring very deep spectroscopy \citep[e.g.][]{Curti2025}, stacking \citep{Isobe2025}, or local universe targets \citep[e.g.][]{MendezDelgado2024} to generate sufficient signal to noise (S/N).  Finally, many iron lines lie in the far-UV or near-IR wavelength ranges, setting upper and lower limits on the available redshifts at which a particular instrument can detect them \citep{Kehoe2025, Stanton2025}. Despite these challenges, we may still study the relative impact of CCSNe and SNe Ia on GCE and stellar populations by using argon as a proxy, even when iron abundances cannot be estimated. 

Argon is created by the $\alpha$-process and thus is expelled with oxygen and other $\alpha$-elements into the ISM during CCSNe \citep{Nomoto2013}. However, white dwarf yields from the models of \citet{Kobayashi2020} suggest $\sim34\%$ of argon enters the ISM via SNe Ia. This modeling indicates that argon in distant star-forming galaxies should be less abundant with respect to oxygen when compared to that of the Sun, consistent with the delayed production of additional argon from SNe Ia. Given that argon is produced alongside oxygen in CCSNe, any deviation in the Ar/O ratio over cosmic time suggests an additional production pathway for argon which does not produce comparable amounts of oxygen \citep{Kobayashi2020a}. The earliest application of argon abundances as a tracer for star formation histories was through planetary nebulae in M31, in which deficient Ar/O suggests a more compact, efficient, and rapid star formation history localized in the thin disk when compared to the MW thin disk and each galaxy's respective thick disks \citep{Arnaboldi2022}.These observations help reveal a bimodality in the $\alpha/\mathrm{Fe}$ ratio within M31's disk, indicating two disparate star-formation events in the past and a radial segmentation of $\alpha-$enhanced stars; These works suggest M31's stellar population evolved in a distinct manner from the MW \citep{Kobayashi2023}. More recently, the imprint of supernovae timescales on the Ar/O ratio has been identified out to $z\sim0.3$ through an analysis of $\sim3000$ Sloan Digital Sky Survey (SDSS) star-forming galaxies, revealing both consistently subsolar Ar/O ratios among the sample as well as a mass-dependent sequence in which more massive galaxies exhibit lower Ar/O at fixed Ar/H, a finding qualitatively consistent with similar studies using the $\alpha/\mathrm{Fe}$ ratio \citep{Bhattacharya2025b}.

Several studies have been carried out to examine this relationship out to Cosmic Noon with mixed results. An analysis of a $z\sim 3$ galaxy as part of the CECILIA survey \citep{Strom2023} constrains its Ar/O abundance ratio to a subsolar value \citep{Rogers2024}. Measurements of 8 galaxies from the JWST EXCELS survey \citep{EXCELS} show similar results, demonstrating the measurement's availability around $z\sim4$ \citep{Stanton2025}. Conversely, an analysis of a lensed $z \simeq 2.37 $ galaxy by \citet{Welch2025} suggests little deviation from the solar value. Another sample of 7 galaxies between $z = 1.3-7.7$ collected by \citet{Bhattacharya2025} suggest that a mix of CCSNe and SNe Ia production pathways enrich the ISM with argon. While intriguing, the small sample size of these studies limits their effectiveness at describing distant star-forming galactic populations as a whole, and may not be representative of the distant star-forming galaxy population overall. To date, the most robust sample of Cosmic Noon \citep[generally defined as $z\sim 1-3$,][]{ForsterSchreiber2020} argon abundances was collected by the CECILIA Collaboration, which reported 36 new measurements of Ar/O \citep{Rogers2025}. This larger sample also reports average subsolar Ar/O, which aligns well with the theoretical considerations outlined above. A representative Ar/O value from a larger sample of galaxies is needed to adequately examine argon's presence in Cosmic Noon nucleosynthetic processes. Such measurements furthermore need to be sensitive enough to detect the relatively faint argon emission lines in the spectra of Cosmic Noon galaxies.

The Assembly of Ultradeep Rest-optical Observations Revealing Astrophysics (AURORA, Program ID: 1914, Co-PIs: Shapley and Sanders) was designed with the express purpose of constructing a representative sample of star-forming galaxies at $z\geq1.3$ from which we may construct empirical calibrations between observed line ratios and direct elemental abundances \citep{Shapley2025, Sanders2025}. The program's primary observing targets were star-forming galaxies at $z\sim2-4$; this range probes the epoch in the Universe during which the global star-formation-rate density was at or near its peak and galactic bulges began to form \citep{Piero2014, Benton2024}. Taking advantage of the unprecedented wavelength range and sensitivity of JWST/NIRSpec, the AURORA program has successfully detected many faint auroral emission lines used to estimate effective temperatures and thus direct abundances of various metals; as such, the AURORA dataset is well suited to detect the faint $[\mbox{ArIII}]\lambda 7137$ feature required for estimating argon abundances through empirical calibrations \citep{Shapley2025, Topping2025, Sanders2025}. Furthermore, the high sampling of galaxies at $z\sim2-4$ permits the construction of representative composite spectra of galaxies at Cosmic Noon, presenting characteristic Ar/O measurements with high S/N. Here we use a sample of 46 star-forming galaxies drawn from the AURORA survey to construct composite spectra in two redshift bins ($z = 2.0-2.6$ and $z=2.8-3.5$). We analyze these composite spectra to determine the average evolution of the Ar/O abundance ratio within the epoch of Cosmic Noon. The size and fairly representative nature of these samples  provides more global measurements compared to previous studies, and thus explores the efficacy of argon as an additional constraint to ISM chemical enrichment pathways during Cosmic Noon.

In Section \ref{section:2}, we discuss both the sample properties and data reduction of the AURORA program, as well as additional data processing of the composite spectra and emission line measurements specific to this analysis. Section \ref{section:3} presents our method for estimating total argon and oxygen abundances through empirical calibrations of line ratios. In Section \ref{section:4} we present our results of representative Ar/O measurements for two redshift bins. In Section \ref{section:5} we compare these results to standard MW chemical evolution models and discuss further applications and future considerations of argon as an additional constraint in both GCE models and observational programs. For this analysis, we adopt the cosmological parameters $H_0 = 70$ $\mathrm{km}$ $ \mathrm{s^{-1}}$ $\mathrm{Mpc^{-1}}$, $\Omega_m = 0.30$, and $\Omega_\Lambda = 0.70$. We also use total oxygen abundance, $12+\log(\mathrm{O/H})$, as a proxy for metallicity, and any reference to ``metallicity" in this analysis should be treated as such. Finally, we adopt Solar abundances, namely $\log(\mbox{Ar/O})_\odot = -2.31$, from \citet{Asplund2021} for consistency with other AURORA projects. We note that more recent versions of these values exist \citep[e.g.,][]{Lodders2025}, but these newer values ultimately do not change our results. 

\begin{figure}[h]
\centering
\includegraphics[width = \columnwidth]{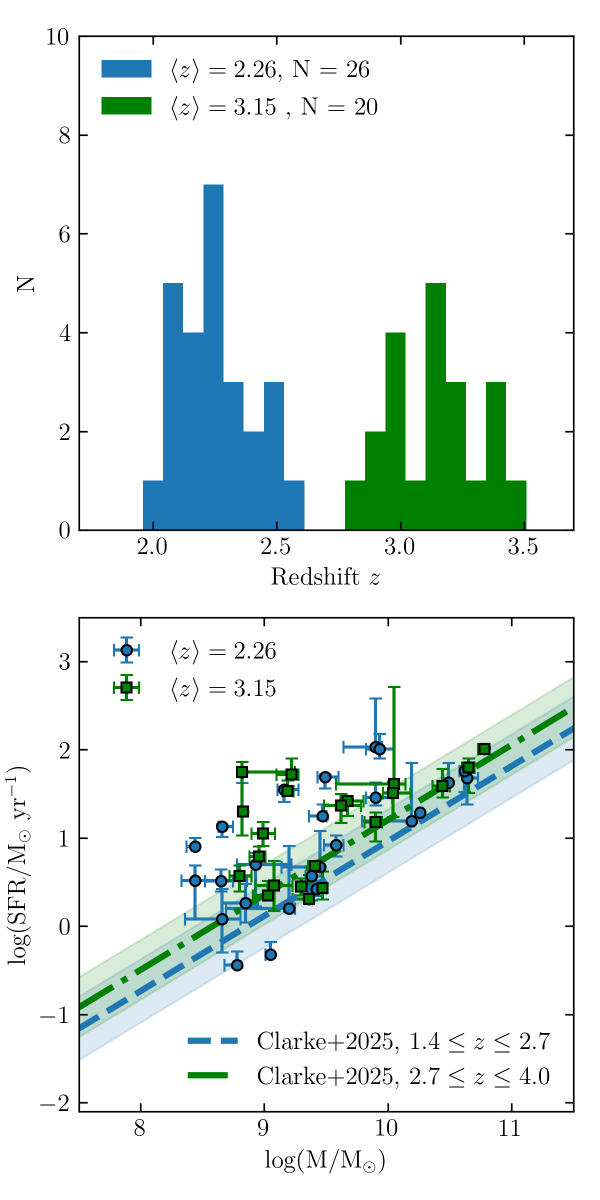}
\caption{Sample properties of the two redshift bins drawn from AURORA. We display the redshift ranges and counts for the two samples (top) and the location of each target on the Star Forming Main Sequence (SFMS, bottom). We include two SFMS evolution lines from \citet{Clarke2025}, corresponding to the $1.4\leq z \leq 2.7$ and $2.7 \leq z \leq 4$ bins included therein. Shading around each line is the $1\sigma$ intrinsic scatter of the corresponding SFMS function \citep{Clarke2025}.}
\label{fig:sampleprops}
\end{figure}

\section{Observations and Methods}\label{section:2}
\subsection{AURORA Observations and Data Reduction}
All spectroscopic data used in this analysis were collected as part of the AURORA Program using the NIRSpec Micro-Shutter Assembly (MSA) on board the James Webb Space Telescope (JWST). A full description of the AURORA observation methods and target selection may be found in previous AURORA works  \citep[e.g.,][]{Sanders2024b, Shapley2025, Topping2025, Sanders2025}. A sample of 97 galaxies was taken from the GOODS-N and COSMOS fields with redshifts $z\geq 1.3$, prioritizing star-forming galaxies at $z \sim 2-4$ where the detection of an auroral emission line ([OIII]$\lambda4363$ or [OII]$\lambda\lambda7320, 7330$) was expected. Additional galaxies were targeted on each of the two AURORA masks including distant galaxies at photometric redshifts $z > 6$, quiescent galaxies at $z > 2$, spectroscopic strong-line emitters at $z > 5$, and galaxies with photometric redshifts of $z > 1.5$. For each mask, the grating/filter combinations G140M/F100LP, G235M/F170LP, and G395M/F290LP were used, with exposure times of 12.3, 8.0, and 4.2 hours, respectively. This strategy yielded resolutions of $R \sim 1000-1500$ over a range of $1 - 5\:\mu\mbox{m}$.  Each observation was collected using a three-point dither pattern at each pointing, and MSA ``slits" were constructed using three microshutters.

Data were reduced following the prescription described in \citet{Sanders2024b}, \citet{Shapley2025}, \citet{Topping2025}, and \citet{Reddy2025}. Raw two-dimensional (2D) spectra from the instrument were corrected for dark current, quantum efficiency bias, detector gain, and cosmic rays. Each 2D spectrum was also calibrated both to a wavelength solution and a photometric reference provided by the Calibration Reference Data System native to JWST. To combine the measured spectra from each dither position, all three spectra were interpolated onto a common grid and combined to form the final 2D spectrum for a particular target. All 2D spectra were converted into  one dimensional (1D) spectra via optimal extraction, and any remaining artifacts were flagged manually. Each 1D spectrum was corrected for slit losses introduced by the small microshutter width, off-center targeting, and the wavelength dependence of the point spread function (PSF) \citep[for a full description of the method, see][]{Reddy2023}. Spectra of the same source in adjacent gratings were calibrated to each other by using spectral features that fell within the overlap range of both gratings. After calibration to each other, grating spectra were scaled to match existing multi-band JWST or HST photometry for an absolute flux calibration. 

Using this available photometry, the spectral energy distribution (SED) was fit using the FAST program \citep{Kriek2009}, assuming the flexible stellar population synthesis models of \citet{Conroy2009} and a \citet{Chabrier2003} IMF. We assume a delayed-$\tau$ star formation history, in which the star-formation rate (SFR) is expressed as $\mathrm{SFR}(t)\propto t\times e^{-t/\tau}$. This expression includes parameters for the time since the onset of star formation, $t$, and the characteristic star-formation timescale, $\tau$. These models were further constrained by assuming a combination of metallicity (in units of solar metallicity, $Z_\odot = 0.014$) and a dust extinction curve in a similar manner to \citet{Reddy2018}; either a model with 1.4 $Z_\odot$ and the dust curve from \citet{Calzetti2000} or a model with 0.27 $Z_\odot$ and the dust curve from \citet{Gordon2003} was chosen via $\chi^2$-minimization. Using this SED estimate as the model for the stellar continuum, we fit each spectral feature with Gaussian profiles to derive their preliminary fluxes. These fluxes were used to correct the photometry for the contributions from nebular emission, and then the corrected SED was re-fit using the above methodology. Using this updated SED model as the underlying continuum, we again fit the emission-line fluxes. These second-pass fluxes were adopted as final for each spectral feature. Since our fluxes are calculated from these SED fits, they include stellar Balmer line absorption contributions by construction. 

Previous AURORA papers have demonstrated that the AURORA dataset is fairly representative of Cosmic Noon galaxies as a whole. The AURORA sample \citep{Shapley2025} exhibits the standard offset toward higher $\mathrm{[NII]\lambda6583/H\alpha}$ and $\mathrm{[OIII]\lambda5008/H\beta}$ \citep[the ``classical" BPT diagram][]{BPT} above the local universe sequence of the same ratios \citep[e.g.,][]{Steidel2014, Shapley2015, Clarke2026}. This is also true of the $\mathrm{[O\:III]\lambda5008/H\beta}$ versus $\mathrm{[S\:II]\lambda\lambda6717, 6731/H\alpha}$ and $\mathrm{[O\:III]\lambda5008/H\beta}$ versus $\mathrm{[O\:I]\lambda6300/H\alpha}$ diagrams, in which the AURORA sample loci are significantly offset from the local universe. The derived masses and SFRs of AURORA galaxies are tightly correlated, suggesting consistency with the star-forming main sequence \citep{Shapley2025, Sanders2025}. Furthermore, as demonstrated in \citet{Topping2025}, the AURORA sample presents electron densities consistent with earlier Cosmic Noon studies \citep[e.g.,][]{Erb2006b, Sanders2016, Davies2021, Reddy2023b}. Finally, a recent analysis by \citet{Khostovan2025} demonstrates that AURORA galaxies with measurable electron temperatures between redshifts $z=1.38-3.5$ exhibit a mass-metallicity relationship (MZR) that is remarkably consistent with previous calibrated MZRs. The depth of AURORA spectra provide high S/N line measurements of the requisite spectral lines in almost all of the sample galaxies, meaning that these galaxies represent the Cosmic Noon galactic population as a whole particularly well. Since these same spectra were used to determine the direct metallicities of much of the AURORA sample in \citet{Sanders2025}, it is likely that these too are fairly representative. 

For this analysis, we selected two subgroups from the AURORA sample: one of 26 galaxies at $z = 2.0-2.6$ (hereafter the $\langle z \rangle = 2.26$ sample) and another of 20 galaxies at $z = 2.8-3.5$ (hereafter the $\langle z \rangle = 3.15$ sample). These two samples were chosen to cover a sizable redshift range while also taking advantage of the high number density of galaxies within this range collected by the AURORA program \citep{Shapley2025}. In this way, we may analyze both the average Ar/O ratio in distant star-forming galaxies and its evolution over cosmic time within the densest portion of the AURORA sample. As shown in Figure~\ref{fig:sampleprops}, these galaxies follow a positive correlation between SFR and stellar mass. Both samples overlap the $z\sim 2-3$ star-forming main sequence but are slightly offset towards higher than average SFR at fixed stellar mass \citep[$+0.40^{+0.60}_{-0.61}$ for the $\langle z \rangle = 2.26$ bin and $+0.23^{+0.54}_{-0.54}$ for the $\langle z \rangle = 3.15$ bin][]{Clarke2025}. We attribute this offset to the AURORA target selection criteria, which prioritized detection of auroral emission lines and thus biases our Cosmic Noon sample toward higher specific star formation rate (sSFR) and lower metallicity at fixed stellar mass. This offset from the star-forming main sequence brings our sample's representativeness into question, as has been noted in other studies of distant star-forming galaxies \citep[e.g.,][]{Scholte2025, Sanders2025}. However, this offset is statistically insignificant within $1\sigma$ confidence, and given the strength of the AURORA dataset more broadly, we may still use this sample to gain some insight into the galactic population and its star-formation histories during Cosmic Noon.

Every galaxy in this sample has coverage of the $[\mbox{Ar\:III}]\lambda 7137$ in the G235M grating, and of the $[\mathrm{O\:II}]\lambda\lambda3727, 3730$ line in the G140M grating. All galaxies within the $\langle z \rangle = 2.26$ stack have coverage of the $\mathrm{[O\:III]\lambda5008}$ in the G140M grating, and galaxies in this sample with $z=2.32-2.60$ (a total of 7 targets) also have coverage in the G235M grating. All galaxies in the $\langle z \rangle = 3.15$ stack have coverage of the same line in the G235M grating exclusively. At this stage, the galaxies selected for this sample only needed \textit{coverage} of these lines, regardless of S/N.  

\subsection{Composite Spectra}\label{section:comp}
Composite spectra can increase the S/N of weaker spectral features, allowing analysis of these weak features for large groups of similar galaxies \citep{Erb2006b, Maiolino2008, Steidel2016, Clarke2024, Sanders2023, Topping2024, Isobe2025}. Given the typical faintness of the $[\mbox{Ar\:III}]\lambda 7137$ emission line, we used median-stacked composite spectra of each redshift bin to improve the S/N of these emission-line measurements. First, each spectrum to be stacked was shifted into its rest frame and interpolated onto a uniform wavelength grid with wavelength points separated by 0.8\AA. This step was performed for each grating for each individual galaxy. When spectra from two adjacent gratings overlapped, the median value of these overlapping points was taken for the final spectrum. These rest-frame spectra were corrected for dust using the following formula:
\begin{equation}
F_{{\rm corr}} = F_{{\rm meas}} \times 10^{0.4E(B-V)k_\lambda}
\label{equation:dustcorrect}
\end{equation}
where $F_{{\rm corr}}$ is the dust-corrected flux at a particular wavelength, $F_{{\rm meas}}$ is the measured flux at that wavelength, $E(B-V)$ is the color excess of the particular galaxy, and $k_\lambda$ is the wavelength-specific dust attenuation coefficient, defined according to \citet{Cardelli1989}. The color excess $E(B-V)$ was calculated using the Balmer decrement from the H$\alpha$, H$\beta$, and H$\gamma$ fluxes of each galaxy when detected, following the procedure described in \citet{Clarke2024b}. In total, the Balmer decrement was calculated using H$\alpha$/H$\beta$ for 42 galaxies, H$\beta$/H$\gamma$ for 3 galaxies, and H$\alpha$/H$\gamma$ for 1 galaxy. Any bad pixels manually flagged during AURORA data reduction were also removed from the final corrected spectra. Finally, each spectrum was normalized by its H$\alpha$ flux, placing all included galaxies onto the same scale, which is ideal for measuring median line ratios. These corrections were applied in an identical matter to the error spectra for each galaxy as well. 

\begin{figure*}[h]
\centering
\includegraphics[width = \textwidth]{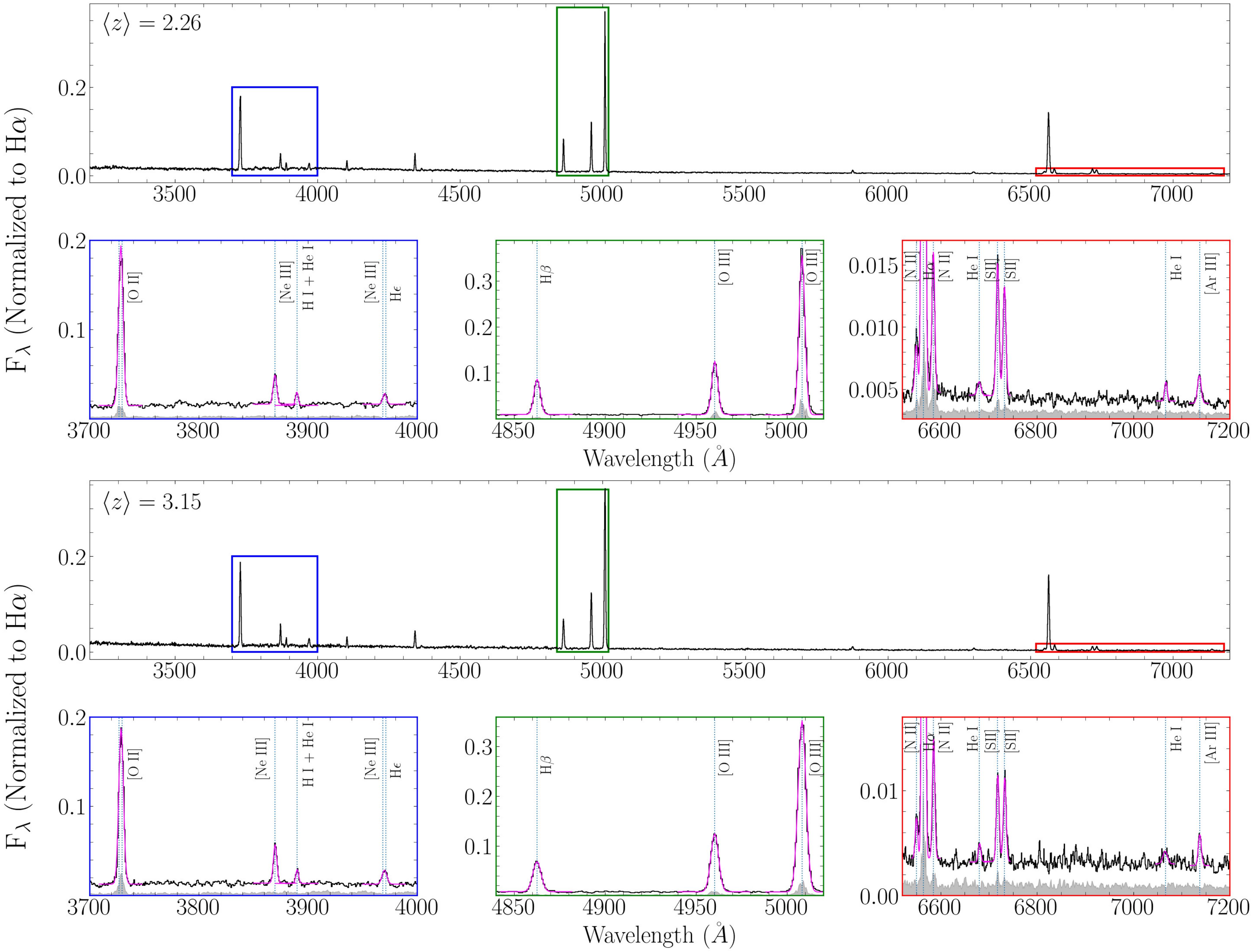}
\caption{Composite Spectra for the $\langle z \rangle = 2.26$ (top panels) and $\langle z \rangle = 3.15$ (bottom panels) redshift bins. Each composite spectrum is normalized to an H$\alpha$ flux of unity and shares the same wavelength grid. The spectra have been trimmed to only show data between 3200 and 8000 \AA. Color-coded insets zoom in on the $[\mathrm{O II}]\lambda\lambda3727, 3730$ and $[\mathrm{Ne\:III}]\lambda3780$ emission lines (blue), the H$\beta$, $\mathrm{[O\:III]}\lambda4960$ and $\mathrm{[O\:III]\lambda5008}$ lines (green), and the H$\alpha$ and $\mathrm{[Ar\:III]}\lambda7137$ lines (red). Gaussian fits to emission features are shown in magenta.}
\label{fig:spectra}
\end{figure*}

The corrected galaxy spectra were then vertically stacked along the uniform wavelength grid, and the median value at each point along this grid was taken as the final composite spectrum value. We note that performing an unweighted mean stack, rather than the median stack described here, does not significantly change our results. Uncertainty on the composite spectra was estimated by bootstrap resampling of the individual spectra with replacement, perturbing each of these resampled spectra by their error, restacking these spectra, and taking the inner 68th percentile boundaries at each point after 1000 iterations. Both composite spectra for the $\langle z \rangle = 2.26$ and $\langle z \rangle = 3.15$ samples are shown in Figure~\ref{fig:spectra}

\subsection{Emission-line Measurements}
In each composite spectrum, the fluxes of each available emission line were measured by fitting Gaussian profiles centered at each spectral feature's rest-frame centroid. Isolated features or blended features unresolved at $R\sim1300$ were fit with single Gaussians, whereas resolved blended features, such as the $[\mathrm{NII}]\lambda\lambda 6550, 6585$ and H$\alpha$ triplet, were fit with the appropriate number of Gaussians simultaneously. The resolution widths at each feature were constrained by taking the median resolution width at that feature from the individual galaxies, and inputting this median into the FWHM as an initial guess. While multi-component fitting of single lines to account for an additional outflow component \citep[see e.g.][]{Schreiber2019} was considered, we found that the multi-component model did not provide a statistically significant improvement over the single-component fit, as indicated by a p-value of $p > 0.01$ from the survival function of the difference in $\chi^2$ statistics \citep{Lam2025}. This test ensures that our models were accurate without overfitting the data. The intrinsic velocity FWHM was inferred from the emission-line FWHM by treating the observed FWHM as the intrinsic and instrumental FWHMs summed in quadrature. As in \citet{Topping2025}, the MSA resolution curves themselves are updated from the JWST user documentation curves based on the ``msafit" package from \citet{deGraaff2024}.

We perform a Monte Carlo simulation with 1000 iterations to calculate confidence intervals for each flux measurement,  perturbing each flux density at each wavelength point and re-fitting Gaussians to the perturbed spectrum. The final confidence interval was taken to be between the upper and lower 68th percentiles from these 1000 iterations, and the $1\sigma$ error for each flux measurement was half the difference between these two values. Every flux measurement that was not greater than 3 times this error was considered undetected and excluded from the following analysis. All of the reported confidence intervals and errors throughout this analysis were carried out in a similar manner. 

\section{Argon and Oxygen Abundances}\label{section:3}
\subsection{Oxygen Abundance}
To determine oxygen abundances from the strong-line ratios present in our composite spectra, we use the empirical calibrations from \citet{Sanders2025}. Following the procedure laid out in \citet{Sanders2021}, we constrain the oxygen abundance with multiple independent line ratio tracers via $\chi^2$ minimization. The inferred oxygen abundance is that which minimizes the following expression:
\begin{equation}
\chi^2=\sum_i\frac{(R_{obs,i}-R_{cal,i}(x))^2}{\sigma_{obs, i}^2+\sigma_{int, i}^2}
\end{equation}
where $R_{obs,i}$ is the observed line ratio for the $i-$th chosen strong-line tracer, $R_{cal,i}(x)$ is the $i-$th calibrated line ratio as a function of the oxygen abundance $x = 12.0+\log(\mathrm{O/H}) - 8.0$, $\sigma_{obs,i}$ is the observed error in the strong-line tracer, and $\sigma_{int, i}$ is the intrinsic scatter of the calibrated line ratio. This intrinsic scatter arises from variations in local gas and ionization conditions at fixed oxygen abundances, which impact the accuracy of empirical calibrations if unaccounted for \citep{Sanders2025}. Following \citet{Sanders2021}, we use the O3, Ne3O2, and O32 ratios as our independent tracers of total oxygen abundance, defined hereafter as
\begin{align}
\mathrm{O3}    &= \frac{[\mathrm{O\:III}]\lambda5008}{\mathrm{H}\beta}\label{Eq:O3}\\
\mathrm{Ne3O2} &= \frac{[\mathrm{Ne\:III}]\lambda3780}{[\mathrm{O\:II}]\lambda\lambda3727,3730}\\
\mathrm{O32}   &= \frac{[\mathrm{O\:III}]\lambda5008}{[\mathrm{O\:II}]\lambda\lambda3727, 3730}
\end{align}
These strong line ratios were also used to determine the oxygen abundances of the individual galaxies for uniformity. Where applicable, these calibrated metallicity values can be compared to directly determined oxygen abundances (see Table \ref{tab:table}). Of the 29 galaxies with $3\sigma$ detections of $\mathrm{[ArIII]}\lambda7137$ and a direct-method oxygen abundance, the median offset between calibrated and direct method abundances is $0.001\pm0.167$ (where the reported error here is the standard deviation of the offsets), suggesting the calibrated values as a whole agree with the direct-method values, and very little systematic offset exists between them. The large differences observed in individual points can be attributed to the large intrinsic scatter inherent to any empirical metallicity calibration. Such conclusions are common among empirical metallicity calibrations for various galactic environments \citep[e.g.,][]{PerezMontero2005, Pilyugin2016, Curti2017}, and this inherent scatter ultimately does not invalidate their ability to describe broad chemical trends in star-forming galaxies.

\subsection{Ar/O Ratio}\label{ArO}
We use a separate empirical calibration to determine Ar/O abundance ratios for our composite spectra and individual galaxies from the emission line ratio Ar3O3, defined as
\begin{equation}
\mathrm{Ar3O3} = \frac{[\mathrm{Ar\:III}]\lambda7137}{\mathrm{[O\:III]\lambda5008}}
\end{equation}
which we adopt throughout the rest of this analysis. 

Generally, including contributions in direct argon abundance calculations from both [Ar III]$\lambda7137$ and [Ar IV]$\lambda\lambda4711,4741$ results in smaller dispersion than just using [Ar III]$\lambda7137$ alone \citep{Arellano-Cordova2024}. While an empirical calibration based on both [Ar III]$\lambda7137$ and [Ar IV]$\lambda\lambda4711,4741$ lines would provide the most robust constraints on total argon abundance, the [Ar IV]$\lambda\lambda4711,4741$ line is often too weak to detect in most spectra \citep[for a few examples of recent measurements of $\mathrm{[Ar IV]}\lambda\lambda4711,4741$ in high-$z$ galaxies, see][]{Rogers2024, Stiavelli2025, Welch2025, Berg2025}. Given that the [Ar IV]$\lambda\lambda4711,4741$ line is not detected in either of our stacks, we can only use the Ar3O3 ratio for our calibration.

It is typically assumed for direct-method oxygen abundances that the total oxygen abundance in star-forming HII regions is traced by the amount of singly- and doubly-ionized oxygen ($\mathrm{O}^+$ and $\mathrm{O}^{++}$), 
with negligible amounts of oxygen present in higher states of ionization \citep{Izotov2006, Berg2021, Sanders2024, Stanton2025}. When the argon abundance is determined directly, it is typically based on measurements of [Ar III]$\lambda7137$ and sometimes additionally [Ar IV]$\lambda\lambda4711,4741$, to which an ionization correction factor (ICF) is  applied to infer the total abundance of argon. The ICF for argon has been shown to depend on the degree of ionization, or, equivalently the ratio of singly ionized oxygen to total oxygen \citep[e.g., $\mathrm{O}^+ / (\mathrm{O}^+ + \mathrm{O}^{++}$][]{Izotov2006}. Here we are using an indirect approach to infer Ar/O. Given the different ICFs for oxygen and argon and their dependencies on the degree of ionization, the Ar3O3 ratio on its own is not an accurate proxy for the total Ar/O ratio. For reference, the ionization potentials of $\mathrm{Ar^{++}}$ and $\mathrm{O^{++}}$ are 27.6 eV and 35.1 eV, meaning a larger fraction of argon atoms are excited into higher ionization states than oxygen for a given stellar radiation field. However, as the O32 line ratio is an effective probe of the degree of ionization, it can be used as an indirect tracer of the ICF. 

\begin{figure}[b]
\centering
\includegraphics[width = \columnwidth]{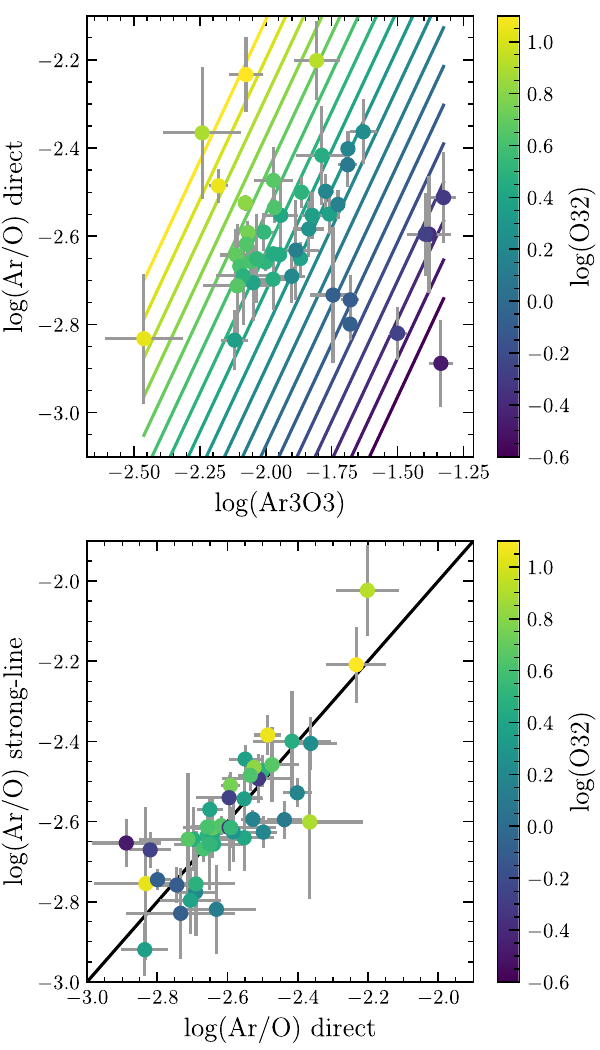}
\caption{Performance of the strong-line calibration of the Ar/O abundance ratio. The top panel compares the line ratio log(Ar3O3) to the direct abundance ratio log(Ar/O). These points are plotted on top of evaluations of Eq. \ref{eq:arCalib} with fixed log(O32), indicated by the color map. The bottom panel plots the log(Ar/O) values of each galaxy with corresponding strong line log(Ar/O) values calculated using Eq. \ref{eq:arCalib}. A color map for log(O32) is again given. The intrinsic scatter about this line is roughly 0.06 dex.}
\label{fig:arCalib}
\end{figure}

This relationship is evident in Figure~\ref{fig:arCalib}, in which direct-method Ar/O ratios of 45 AURORA and literature galaxies (R. Sanders et al. 2025, in prep.) are correlated with the Ar3O3 ratio and anti-correlated with the O32 ratio. We therefore find that a 2D function of both Ar3O3 and O32 can adequately reproduce direct-method Ar/O abundances. Fitting a function linear in both log(Ar3O3) and log(O32) to these directly-determined Ar/O data points yields  
\begin{align}
\log(\mathrm{Ar/O}) 
   &= -0.523 
      + 1.277 \log(\mathrm{Ar3O3}) \nonumber \\
   &\quad + 0.878 \log(\mathrm{O32})
   \label{eq:arCalib}
\end{align}
with an intrinsic scatter (after accounting for measurement uncertainty) of roughly 0.06 dex (see Figure~\ref{fig:arCalib}).  We use this calibration to determine the Ar/O abundances of our composite spectra and individual galaxies, convolving this 0.06 dex scatter into the final confidence intervals. We opt to use these calibrations instead of direct-method calculations to demonstrate their efficacy for galaxies without significant detection of auroral lines needed to estimate temperature. Furthermore, given that significant detections of auroral lines were not a requirement for inclusion in our sample, the strong-line calibrations are used for consistency. A more robust discussion of direct Ar/O measurements will be found in Sanders et al. 2026, in prep.

Besides standard observational dispersion, there are several sources of systematic uncertainty which contribute to the scatter about Eq. \ref{eq:arCalib} that arise from our particular assumptions about the dust and ionization conditions of our sample and the AURORA dataset more broadly. Firstly, several analyses \citep[e.g.,][]{Sanders2024b, Reddy2025} have demonstrated that the dust attenuation curve from \citet{Cardelli1989} may not adequately capture the opacity of the AURORA sample. While we adopt the \citet{Cardelli1989} law to align our results with the rest of the literature, these differences between the assumed and actual dust attentuation may generate additional scatter about the calibration shown in Figure~\ref{fig:arCalib}. To study this discrepancy, we also recompute the fluxes and abundances used to generate Eq.~\ref{eq:arCalib} and Figure~\ref{fig:arCalib} assuming the dust attenuation law from \citet{Reddy2025}; this law uses the AURORA sample's hydrogen recombination lines to explore the highly reddened environments observed in AURORA, and proposes a sub-unity dust covering fraction to explain the observed differences between the \citet{Cardelli1989} law and the observed nebular attenuation \citep{Reddy2025}. After recalculating the direct Ar/O abundances among the individual sources, the calibration becomes
\begin{align}
\log(\mathrm{Ar/O})_{\mathrm{R26}} 
   &= -0.487 
      + 1.267 \log(\mathrm{Ar3O3}_{\mathrm{R26}}) \nonumber \\
   &\quad + 0.750 \log(\mathrm{O32_{\mathrm{R26}}})
   \label{eq:ReddyCalib}
\end{align}
where ``R26" denotes usage of the dust attenuation coefficients from \citet{Reddy2025} to dust-correct the appropriate fluxes. This calibration has similar intrinsic scatter to that of Eq.~\ref{eq:arCalib}, at 0.07 dex. The largest change comes from the O32 term, since the O32 ratio is more sensitive to dust attenuation assumptions than Ar3O3, which changes very little. We note that these new coefficient values in Eq.~\ref{eq:ReddyCalib} are well within the $1\sigma$ Monte Carlo error of the original coefficients in Eq.~\ref{eq:arCalib}. Furthermore, if we instead examine how much the predicted $\log(\mathrm{Ar/O})$ value changes based on these coefficient uncertainties across the calibration sample's Ar3O3 and O32 ratios, we find an average scatter of 0.06 dex, comparable to the reported intrinsic scatter. Using these coefficients and the same two stacks (constructed using the same process outlined in Section \ref{section:2} but replacing the \citet{Cardelli1989} law with that from \citet{Reddy2025}), we find that the resulting Ar/O abundance ratios remain within $1\sigma$ of the ratios reported below. This result suggests that variations among the chosen dust attenuation curve do not dramatically impact our results, but still result in additional scatter inherent to Eq. \ref{eq:arCalib}.

Another source of systematic uncertainty is our assumptions about the ionization structure of these distant galaxies. In many direct-method abundance studies \citep[including][]{Sanders2025}, calculations are performed assuming single-zone, low density ionization conditions. In other words, while different high- and low-ionization temperatures are assumed, only one such density is used to determine ionic abundances, and thus full elemental abundances later on. While this assumption is oftentimes sufficient, new analysis suggests that the single-zone density assumptions may lead to overestimated electron temperatures, and thus underestimated ionic and elemental abundances \citep{Martinez2025}. In particular, high density ($n_e\sim10^5\:\mathrm{cm^{-3}}$) environments experience greater degrees of collisional de-excitation, suppressing forbidden transition line strengths and thus biasing the line ratios used to compute and express direct abundances \citep{Martinez2025}. As a result of poor density assumptions, electron temperatures are overestimated and the resulting abundances are underestimated. This effect is especially important for our case, in which both the O32 ratio and the direct Ar/O ratio (through its ICF dependence on the relative abundances of $\mathrm{O^+}$ and $\mathrm{O^{++}}$) are heavily dependent on ionization conditions, and thus proper assumptions about the density structure of our targets as well. While \citet{Martinez2025} suggests a multi-zone density model for better accuracy, we adhere to the one-zone model in this work for uniformity with previous AURORA studies; this naive assumption may also explain the scatter observed about the 1:1 line in Figure~\ref{fig:arCalib}. However, we note that Cosmic Noon galaxies, both in our sample and more generally \citep[e.g.,][]{Sanders2016, Kashino2017, Maseda2017, Topping2025}, typically do not have electron densities of order greater than $10^{4}\:\mathrm{cm}^{-3}$; thus any systematic uncertainty from poor density assumptions must be minimal, further justifying our decision to use a one-zone density model.

\section{Results}\label{section:4}
\subsection{Composite Spectra}
By using the calibrations described in Section \ref{section:3}, we find oxygen abundances of $12 + \log(\mbox{O/H}) = 8.32^{+0.04}_{-0.04}$ for the $\langle z \rangle = 2.26$ sample and $12 + \log(\mbox{O/H}) = 8.25^{+0.04}_{-0.04}$ for the $\langle z \rangle = 3.15$ sample. These values are both subsolar and consistent with the MZR observed at $z\sim1-4$ in which metallicity decreases with increasing redshift at fixed $M_*$ \citep{Sanders2015, Sanders2021, Strom2022, Papovich2022, Sanders2024, Stanton2025b}. The $z\sim2.3$ and $z\sim3.1$ MZR from \citet{Sanders2021} predicts a $12+\log(\mathrm{O/H})= 8.37^{+0.03}_{-0.03}$ and $12+\log(\mathrm{O/H})= 8.31^{+0.04}_{-0.04}$ at the median masses of the $\langle z \rangle = 2.26$ and $\langle z \rangle = 3.15$ samples, agreeing within one standard deviation with our calibration results.

\begin{figure*}[h]
\centering
\includegraphics[width = \textwidth]{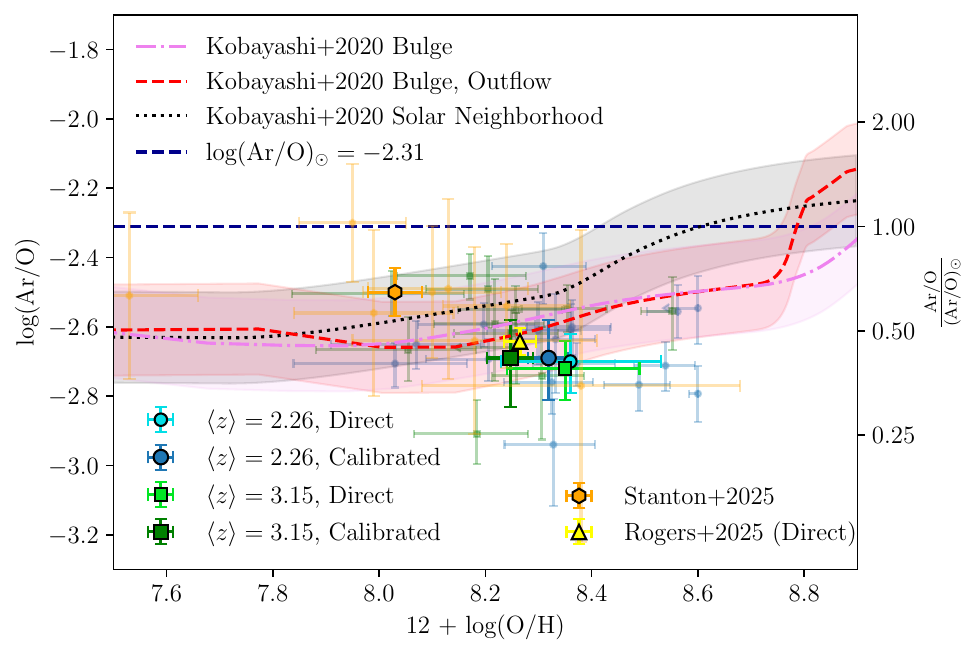}
\caption{Evolution of the Ar/O abundance ratio for the two redshift samples, with the $\langle z \rangle = 2.26$ sample as the blue circle and the $\langle z \rangle = 3.15$ sample as the green square. Direct abundances for the same stacks (see Section \ref{subsection:direct}) are shown as the light blue circle and green square. included in this plot are the dashed dark blue line and marker for the solar Ar/O ratio value ($\log(\mbox{Ar/O})_\odot = -2.31$) from \citet{Asplund2021}, the centroid and individual points (in orange) from \citet{Stanton2025}, and the inverse-variance weighted average Ar/O ratio of 19 galaxies from \citet{Rogers2025} galaxies with direct oxygen abundances (yellow). Here we also include three GCE models from \citet{Kobayashi2020}, namely the Galactic Bulge with onset of galactic winds at $t = 3 \:\:\mathrm{Gyr}$ (violet), Galactic Bulge with outflow (red), and the Solar Neighborhood (black) models. The shaded regions surrounding the model curves are based on  the uncertainties in the solar abundances from \citet{Asplund2010}. Points with arrows for error bars indicate galaxies at the upper limit of the metallicity calibration \citep[$12+\log(\mathrm{O/H}) = 8.60$,][]{Sanders2025}.}
\label{fig:moneyplot}
\end{figure*}

Additionally, we find Ar/O abundance ratios of $-2.69^{+0.11}_{-0.12}$ for the $\langle z \rangle = 2.26$ sample, and $-2.69^{+0.11}_{-0.13}$ for the $\langle z \rangle = 3.15$ sample. In terms of linearized solar abundances, these values are $0.42^{+0.12}_{ -0.10}\mbox{(Ar/O)}_{\odot}$ for the $\langle z \rangle = 2.26$ sample and $0.42^{+ 0.12}_{-0.11}\mbox{(Ar/O)}_{\odot}$ for the $\langle z \rangle = 3.15$ sample. These results, along with the solar value $\log(\mbox{Ar/O})_\odot=-2.31$, the GCE models in Section \ref{section:5}, and some additional results from the literature for context, are summarized in Figure \ref{fig:moneyplot}. We also include the line ratios used in both calibrations in Table \ref{tab:table}.

\begin{deluxetable*}{@{\hskip 12pt}lccccccccccc@{\hskip 12pt}}
\setlength{\tabcolsep}{2pt}
\tabletypesize{\scriptsize} 
\tablecaption{Detected sample properties and line ratios \label{tab:table}}
\tablehead{
\colhead{ID} & \colhead{$z$} & \colhead{$\log\left(\frac{M}{M_\odot}\right)$} &
\colhead{$\log\left(\mathrm{\frac{SFR}{\mathrm{M}_\odot\,\mathrm{yr}^{-1}}}\right)$} &
\colhead{$\log$(Ar3O3)} & \colhead{$\log$(O32)} & \colhead{$\log$(Ne3O2)} &
\colhead{$\log$(O3)} & \colhead{$\log(\mathrm{Ar/O})$} &
\colhead{$12+\log(\mathrm{O/H})_{\mathrm{cal}}$} &
\colhead{$12+\log(\mathrm{O/H})_{\mathrm{dir}}$}
}
\startdata
&&&&&&&&&\\
COSMOS-3324   & 2.308 & $10.64^{+0.09}_{-0.00}$ & $2.087^{+0.033}_{-0.031}$ & $-1.33^{+0.05}_{-0.06}$ & $-0.370^{+0.011}_{-0.010}$  & $-1.593^{+0.116}_{-0.107}$ & $0.218^{+0.014}_{-0.014}$  & $-2.55^{+0.09}_{-0.10}$ & $8.60^{+0.00}_{-0.05}$  & $8.30^{+0.22}_{-0.20}$ \\
COSMOS-4029   & 2.076 & $8.44^{+0.00}_{-0.07}$  & $1.227^{+0.008}_{-0.009}$ & $-2.09^{+0.03}_{-0.02}$ & $0.614^{+0.003}_{-0.003}$  & $-0.451^{+0.005}_{-0.005}$ & $0.836^{+0.003}_{-0.003}$  & $-2.65^{+0.06}_{-0.07}$ & $8.07^{+0.12}_{-0.16}$  & $8.21^{+0.03}_{-0.03}$ \\
COSMOS-4113   & 3.085 & $10.40^{+0.06}_{-0.08}$ & $1.287^{+0.023}_{-0.024}$ & $-1.43^{+0.07}_{-0.07}$ & $-0.232^{+0.008}_{-0.008}$ & $-1.312^{+0.055}_{-0.056}$ & $0.270^{+0.011}_{-0.011}$  & $-2.55^{+0.10}_{-0.11}$ &$ 8.55^{+0.00}_{-0.06}$ & \nodata \\
COSMOS-4156   & 2.190 & $8.66^{+0.09}_{-0.03}$  & $1.253^{+0.007}_{-0.007}$ & $-2.19^{+0.02}_{-0.02}$ & $0.692^{+0.003}_{-0.003}$  & $-0.376^{+0.005}_{-0.005}$ & $0.807^{+0.002}_{-0.002}$  & $-2.71^{+0.06}_{-0.06}$ & $8.03^{+0.14}_{-0.19}$  & $8.08^{+0.06}_{-0.06}$ \\
COSMOS-4429   & 2.102 & $8.93^{+0.29}_{-0.1}$  & $0.481^{+0.023}_{-0.023}$ & $-1.64^{+0.06}_{-0.07}$ & $0.223^{+0.007}_{-0.007}$  & $-0.727^{+0.020}_{-0.020}$ &$ 0.614^{+0.007}_{-0.008}$  & $-2.43^{+0.10}_{-0.10}$ & $8.31^{+0.08}_{-0.10}$ & $8.07^{+0.11}_{-0.09}$ \\
COSMOS-4740   & 3.155 & $10.05^{+0.09}_{-0.47}$ & $1.927^{+0.010}_{-0.011}$ & $-1.70^{+0.03}_{-0.03}$ & $0.165^{+0.004}_{-0.004}$  & $-0.845^{+0.015}_{-0.015}$ & $0.568^{+0.004}_{-0.004}$  & $-2.55^{+0.07}_{-0.08}$ & $8.35^{+0.07}_{-0.09}$ & $8.29^{+0.10}_{-0.08}$ \\
COSMOS-5283   & 2.174 & $9.47^{+0.03}_{-0.11}$  & $1.573^{+0.006}_{-0.006}$ & $-1.99^{+0.01}_{-0.01}$ &$ 0.418^{+0.002}_{-0.002}$  & $-0.685^{+0.005}_{-0.004}$ & $0.760^{+0.002}_{-0.002}$  & $-2.69^{+0.07}_{-0.06}$ & $8.21^{+0.10}_{-0.12}$ & $8.27^{+0.03}_{-0.02}$ \\
COSMOS-5571   & 2.278 & $10.26^{+0.02}_{-0.01}$ & $1.414^{+0.012}_{-0.012}$ & $-1.68^{+0.03}_{-0.03}$ & $-0.114^{+0.004}_{-0.004}$ & $-1.179^{+0.023}_{-0.024}$ & $0.392^{+0.005}_{-0.005}$  & $-2.77^{+0.07}_{-0.07}$ & $8.49^{+0.06}_{-0.07}$ & $8.46^{+0.10}_{-0.10}$ \\
COSMOS-5901   & 2.397 & $9.90^{+0.13}_{-0.08}$  & $1.239^{+0.019}_{-0.016}$ & $-1.41^{+0.04}_{-0.05}$ & $-0.269^{+0.010}_{-0.010}$ & $-1.314^{+0.052}_{-0.059}$ & $0.249^{+0.012}_{-0.012}$  & $-2.56^{+0.08}_{-0.08}$ & $8.56^{+0.00}_{-0.06}$ & $8.39^{+0.19}_{-0.18}$ \\
COSMOS-7883   & 2.153 & $9.45^{+0.02}_{-0.31}$  & $0.876^{+0.003}_{-0.003}$ & $-1.83^{+0.05}_{-0.07}$& $0.277^{+0.004}_{-0.004}$  & $-0.842^{+0.014}_{-0.015}$ & $0.663^{+0.006}_{-0.006}$  & $-2.61^{+0.09}_{-0.10}$ & $8.30^{+0.08}_{-0.07}$ & $8.10^{+0.10}_{-0.06}$ \\
COSMOS-8363   & 3.248 & $9.62^{+0.16}_{-0.14}$  & $1.573^{+0.014}_{-0.014}$ & $-1.89^{+0.03}_{-0.03}$ & $0.359^{+0.003}_{-0.003}$  & $-0.737^{+0.009}_{-0.010}$ &$ 0.709^{+0.005}_{-0.005}$  & $-2.62^{+0.07}_{-0.07}$ & $8.25^{+0.09}_{-0.11}$ & $8.39^{+0.06}_{-0.05}$ \\
COSMOS-8697   & 3.207 & $8.82^{+0.43}_{-0.01}$  & $0.990^{+0.031}_{-0.035}$ & $-1.923^{+0.09}_{-0.12}$ & $0.442^{+0.007}_{-0.006}$  & $-0.744^{+0.030}_{-0.030}$ &$ 0.751^{+0.014}_{-0.015}$  & $-2.59^{+0.13}_{-0.15}$ & $8.22^{+0.09}_{-0.12}$ & \nodata \\
GOODSN-11584  & 3.362 & $10.78^{+0.00}_{-0.05}$ & $2.486^{+0.019}_{-0.018}$ & $-2.13^{+0.05}_{-0.06}$ & $0.373^{+0.003}_{-0.003}$  & $-0.662^{+0.011}_{-0.010}$ & $0.824^{+0.006}_{-0.007}$  & $-2.91^{+0.09}_{-0.09}$ & $8.18^{+0.10}_{-0.12}$ & $8.52^{+0.17}_{-0.09}$ \\
GOODSN-19067  & 2.281 & $9.58^{+0.06}_{-0.10}$  & $0.894^{+0.015}_{-0.016}$ & $-1.69^{+0.04}_{-0.04}$ &$ 0.093^{+0.005}_{-0.005}$& $-0.867^{+0.015}_{-0.016}$ & $0.575^{+0.006}_{-0.006}$ & $-2.60^{+0.08}_{-0.08}$ & $8.36^{+0.07}_{-0.08}$ & $8.16^{+0.11}_{-0.08}$ \\
GOODSN-19848  & 2.992 & $9.19^{+0.05}_{-0.04}$  & $1.523^{+0.012}_{-0.013}$ & $-1.86^{+0.02}_{-0.03}$ & $0.512^{+0.004}_{-0.004}$  & $-0.547^{+0.009}_{-0.009}$ & $0.720^{+0.005}_{-0.005}$  & -$2.45^{+0.07}_{-0.07}$ & $8.17^{+0.11}_{-0.14}$ & $8.35^{+0.09}_{-0.08}$ \\
GOODSN-21033  & 3.112 & $8.96^{+0.05}_{-0.08}$  & $1.225^{+0.010}_{-0.010}$ & $-2.06^{+0.02}_{-0.02}$ & $0.735^{+0.003}_{-0.003}$  & $-0.413^{+0.006}_{-0.006}$ & $0.825^{+0.004}_{-0.004}$ & $-2.51^{+0.06}_{-0.07}$ & $8.02^{+0.14}_{-0.19}$ &$ 8.07^{+0.02}_{-0.02}$ \\
GOODSN-21522  & 2.363 & $9.49^{+0.11}_{-0.06}$  & $1.081^{+0.011}_{-0.010}$ & $-1.76^{+0.03}_{-0.03}$& $0.189^{+0.003}_{-0.003}$  & $-0.890^{+0.011}_{-0.013}$ & $0.558^{+0.004}_{-0.004}$  & $-2.61^{+0.07}_{-0.07}$ & $8.36^{+0.07}_{-0.09}$ & $8.10^{+0.10}_{-0.07}$ \\
GOODSN-21726  & 3.409 & $8.83^{+0.00}_{-0.00}$  & $1.638^{+0.054}_{-0.047}$ & $-1.83^{+0.10}_{-0.12}$ & $0.136^{+0.010}_{-0.009}$  & $-0.821^{+0.077}_{-0.070}$ & $0.690^{+0.018}_{-0.019}$  & $-2.74^{+0.14}_{-0.17}$ & $8.31^{+0.08}_{-0.09}$ & \nodata \\ 
GOODSN-22235  & 2.430 & $9.16^{+0.12}_{-0.05}$  & $1.446^{+0.012}_{-0.012}$ & $-1.93^{+0.01}_{-0.01}$ & $0.443^{+0.002}_{-0.002}$  & $-0.632^{+0.005}_{-0.005}$ & $0.743^{+0.004}_{-0.004}$  & $-2.59^{+0.06}_{-0.06}$ & $8.20^{+0.10}_{-0.12}$ & $8.10^{+0.10}_{-0.07}$ \\
GOODSN-22932  & 3.331 & $8.99^{+0.10}_{-0.09}$  & $1.182^{+0.014}_{-0.014}$ & $-2.13^{+0.05}_{-0.06}$ &$ 0.652^{+0.005}_{-0.004}$  & $-0.415^{+0.008}_{-0.009}$ & $0.814^{+0.005}_{-0.005}$  &$ -2.66^{+0.09}_{-0.10}$ & $8.06^{+0.13}_{-0.17}$ & $8.09^{+0.05}_{-0.05}$ \\
GOODSN-23927  & 3.364 &$ 9.41^{+0.03}_{-0.06}$  & $0.885^{+0.018}_{-0.019}$ & $-1.83^{+0.06}_{-0.08}$ & $0.229^{+0.006}_{-0.005}$& $-0.700^{+0.015}_{-0.015}$ & $0.722^{+0.007}_{-0.007}$  & $-2.66^{+0.10}_{-0.11}$ &$ 8.26^{+0.09}_{-0.10}$ & \nodata \\
GOODSN-26798  & 2.483 & $10.49^{+0.00}_{-0.04}$ & $1.576^{+0.012}_{-0.012}$& $-1.52^{+0.03}_{-0.04}$ & $-0.283^{+0.004}_{-0.004}$ & $-1.138^{+0.028}_{-0.025}$ & $0.274^{+0.005}_{-0.006}$  & $-2.71^{+0.107}_{-0.07}$ & $8.54^{+0.06}_{-0.06} $& $8.62^{+0.10}_{-0.10}$ \\
GOODSN-27876  & 2.271 & $10.19^{+0.02}_{-0.40}$ & $1.618^{+0.016}_{-0.016}$ & $-1.41^{+0.05}_{-0.05}$ & $-0.540^{+0.008}_{-0.008}$ & $-1.537^{+0.082}_{-0.088}$ & $-0.006^{+0.009}_{-0.008}$ & $-2.79^{+0.08}_{-0.09}$ & $8.60^{+0.00}_{-0.02}$ & $8.65^{+0.10}_{-0.10}$ \\
GOODSN-28209  & 3.233 & $9.22^{+0.06}_{-0.12}$  & $1.908^{+0.014}_{-0.013}$ & $-1.83^{+0.02}_{-0.03}$ & $0.355^{+0.004}_{-0.004}$  & $-0.677^{+0.011}_{-0.014}$ & $0.664^{+0.005}_{-0.005}$  & $-2.55^{+0.06}_{-0.07}$ & $8.26^{+0.09}_{-0.11}$ & $8.52^{+0.13}_{-0.08}$ \\
GOODSN-30053  & 2.245 & $9.90^{+0.05}_{-0.26}$ & $1.867^{+0.010}_{-0.010}$ & $-1.78^{+0.02}_{-0.02}$ &$ 0.103^{+0.003}_{-0.003}$  & $-0.856^{+0.011}_{-0.012}$ & $0.547^{-0.004}_{+0.004}$  & $-2.71^{+0.06}_{-0.07}$ & $8.37^{+0.07}_{-0.08}$ & $8.31^{+0.07}_{-0.05}$ \\
GOODSN-30811  & 2.307 & $9.42^{+0.02}_{-0.04}$  & $0.730^{+0.015}_{-0.016}$ &$ -1.88^{+0.05}_{-0.05}$ &$ 0.182^{+0.004}_{-0.004}$ & $-0.867^{+0.013}_{-0.014}$ & $0.644^{+0.007}_{-0.006}$  & $-2.76^{+0.09}_{-0.09}$ & $8.33^{+0.08}_{-0.09}$ & $8.32^{+0.13}_{-0.10}$ \\
GOODSN-917938 & 2.922 & $8.80^{+0.10}_{-0.08}$  & $0.418^{+0.036}_{-0.044}$ & $-1.77^{+0.07}_{-0.07}$ & $0.335^{+0.008}_{-0.007}$  & $-0.653^{+0.019}_{-0.021}$ & $0.784^{+0.016}_{-0.016}$& $-2.49^{+0.09}_{-0.11}$ & $8.21^{+0.10}_{-0.12}$ & \nodata \\
GOODSN-919329 & 2.575 & $8.85^{+0.10}_{-0.19}$  & $0.289^{+0.033}_{-0.036}$ & $-1.73^{+0.10}_{-0.12}$ & $0.103^{+0.008}_{-0.008}$  & $-0.985^{+0.036}_{-0.037}$ & $0.696^{+0.014}_{-0.013}$& $-2.64^{+0.12}_{-0.17}$ & $8.33^{+0.08}_{-0.09}$ & \nodata \\
GOODSN-931951 & 2.483 & $8.65^{+0.10}_{-0.13}$  & $0.566^{+0.019}_{-0.017}$ & $-2.01^{+0.11}_{-0.12}$ & $0.169^{+0.005}_{-0.006}$  & $-0.760^{+0.015}_{-0.016}$ & $0.598^{+0.007}_{-0.006}$& $-2.94^{+0.14}_{-0.19}$ & $8.33^{+0.08}_{-0.09}$ & \nodata \\
\cutinhead{Stacks}
$\langle z \rangle = 2.26$ Stack & 2.255 & $9.54^{+0.07}_{-0.10}$ & $1.458^{+0.015}_{-0.015}$ &
$-1.82^{+0.04}_{-0.04}$ & $0.179^{+0.006}_{-0.005}$ & $-0.832^{+0.014}_{-0.021}$ &
$0.631^{+0.011}_{-0.011}$ & $-2.69^{+0.11}_{-0.12}$ & $8.32^{+0.03}_{-0.04}$ & $8.36^{+0.17}_{-0.13}$ \\
$\langle z \rangle = 3.15$ Stack & 3.145 & $9.64^{+0.08}_{-0.14}$ & $1.191^{+0.015}_{-0.016}$ &
$-1.90^{+0.08}_{-0.10}$ & $0.299^{+0.007}_{-0.007}$ & $-0.739^{+0.048}_{-0.044}$ &
$0.734^{+0.024}_{-0.026}$ & $-2.69^{+0.11}_{-0.14}$ & $8.25^{+0.04}_{-0.04}$ & ${8.35^{+0.14}_{-0.11}}$ \\ \\
\enddata
\tablecomments{Blank entries indicate galaxies/stacks with no direct method oxygen abundances. $12+\log(\mathrm{O/H})_{cal}$ are the calibrated oxygen abundances, and $12+\log(\mathrm{O/H})_{dir}$ are the direct method oxygen abundances from either \citet{Sanders2025} (individual galaxies) or this work (composite spectra, \ref{subsection:direct}).}
\end{deluxetable*}

\subsection{Individual Galaxies}
Of the 46 galaxies included the composite spectra, 29 had $3\sigma$ detections of $[\mathrm{Ar\:III}]\lambda7137$, meaning the same calibrations from Section \ref{section:3} could be applied to determine their Ar/O abundance ratios. In total, Ar/O abundances were estimated for 17 galaxies from the $\langle z \rangle = 2.26$ stack and 12 from the $\langle z \rangle = 3.15$ stack.
 We calculate oxygen abundances following the method used for the composite spectra as well. The results for individual galaxies are summarized in Table~\ref{tab:table}, along with the Ar3O3, O3, Ne3O2, and O32 ratios, their masses, and SFRs. These galaxies are also displayed alongside the composite spectra in Figure~\ref{fig:moneyplot}.

\section{Discussion}\label{section:5}

\subsection{Comparison between calibrated and direct Ar/O}\label{subsection:direct}
While it is clear that, by construction, the Ar/O calibration (Eq. \ref{eq:arCalib}) adequately reproduces direct Ar/O abundances for the individual galaxies, it is yet unclear if the calibration performs equally well for the composite spectra. Since both composite spectra have significant detections of $\mathrm{[O\:III]\lambda 4363}$, we may constrain each composite spectrum's electron temperature, $T_e$, and from them derive their respective direct Ar/O abundances \citep{Searle1972, Sanders2024, Stanton2025, Sanders2025}. We compute the temperature, density, and necessary ionic abundances ($\mathrm{O^{++}/H^{++}},\: \mathrm{O^+/H^+}, \:\mathrm{and} \:\mathrm{Ar^{++}/H^{+}}$) for each composite spectrum using the Python package \texttt{pyneb} \citep[v1.1.28,][]{Luridiana2015}. Note that the fluxes used during this portion are already dust corrected, as described in Section \ref{section:comp}. As is standard for \texttt{pyneb}, our fluxes were renormalized to their respective $\mathrm{H\beta}$ fluxes instead of $\mathrm{H\alpha}$, as was done during the stacking (see Section~ \ref{section:2}).

In general, we follow the prescriptions for direct abundances outlined in \citet{Sanders2025}. We derive electron density, $n_e(\mathrm{S^+})$ using the $\mathrm{[S\:II]\lambda6718}/\mathrm{[S\:II]\lambda6733}$ flux ratio. For the $\mathrm{S^+}$ ion, we adopt transmission probabilities from \citet{Rynkun2019} and collision strengths from \citet{Tayal2010}. We derive the high-ionization temperature, $T_e(\mathrm{O^{++}})$ using the 
$\mathrm{([O\:III]\lambda4959 +[O\:III]\lambda5007) /[O\:III]\lambda4363}$ flux ratio. For the $\mathrm{O^{++}}$ ion, we adopt transmission probabilities from \citet{Froese2004} and collision strengths from \citet{Storey2014}. Since temperature and density are dependent on each other, we must solve for them iteratively; by first supplying an initial guess temperature ($T_e =10,000$ K), we solve for $n_e(\mathrm{S^+})$, then solve for $T_e(\mathrm{O^{++}})$ using this new density. This process was repeated until both temperature and density converged, with the convergence criteria set to a $>1\%$ difference between successive iterations of $n_e(\mathrm{S^+})$ and a $>0.5\%$ difference between successive iterations of $T_e(\mathrm{O^{++}})$. 

Ideally, the low-ionization temperature ($T_e(\mathrm{O^{+}})$) would be derived using the $\mathrm{[O\:II]\lambda\lambda7322,\:7332/}$
$\mathrm{[O\:II]\lambda\lambda3727,\:3730}$ flux ratio \citep{Sanders2025}. However, the $\mathrm{[O\:II]\lambda7322}$ line is not significantly detected in the $\langle z\rangle = 3.15$ stack meaning we cannot constrain $T_e(\mathrm{O^{+}})$ effectively using the above ratio. Instead, we opt to use the scaling relationship from \citet{Campbell1986}:
\begin{equation}
T_e(\mathrm{O^+}) = 0.7\times T_e(\mathrm{O^{++})} + 3000\:\mathrm{K}
\label{equation:temp}
\end{equation}
It has been demonstrated that temperature scaling relationships like Eq. \ref{equation:temp} are more uncertain at high redshift \citep{ArellanoCordova2020, Rogers2022}. However, since the $\langle z\rangle = 3.15$ stack is missing the $\mathrm{[O\:II]\lambda7322}$ line \citep[as well as the $\mathrm{[S\:II]\lambda4070}$, which can also constrain the low-ionization temperature,][]{Sanders2025}, we are left with little choice. For uniformity, we use Eq. \ref{equation:temp} to calculate $T_e(\mathrm{O^+})$ for the $\langle z\rangle = 2.26$ stack as well. 

Using the derived densities and temperatures, we use $\texttt{pyneb}$ to compute $\mathrm{O^{++}/H^{++}},\: \mathrm{O^+/H^+}, \:\mathrm{and} \:\mathrm{Ar^{++}/H^{+}}$. $\mathrm{O^{++}/H^{++}}$ was calculated using the O3 ratio (Eq. \ref{Eq:O3}) and the above $\mathrm{O^{++}}$ transition probabilities and collision strengths. Simiarly, $\mathrm{O^+}$ was calculated using the $\mathrm{[O\:II]\lambda\lambda3727,\:3730} /\mathrm{H\beta}$ flux ratio, the transition probabilities from \citet{Zeippen1982} (up to level 5) and \citet{Wiese1996} (6+), and the collision strengths from \citet{Kisielius2009}. For $\mathrm{Ar^{++}/H^{+}}$, we use the $\mathrm{[Ar\:III]\lambda7137}$ line flux, and both the transition probabilities and collision strengths from \citet{Munoz2009}.

We compute the O/H abundance using the assumptions outlined in Section \ref{ArO} (i.e., summing $\mathrm{O^{++}/H^{++}}$ and $\mathrm{O^+/H^+}$ and assuming minimal contributions from higher ionization states and neutral O). Following the  discussion in Section~\ref{ArO} we convert the $\mathrm{Ar^{++}/H^{+}}$ into the Ar/H abundance ratio using the ICF from \citet{Izotov2006}. 
Using the procedure outlined above, we calculate direct $\mathrm{12+\log(O/H)}$ and $\mathrm{log(Ar/O)}$ values of $8.36^{+0.17}_{-0.13}$ and $-2.70^{+0.08}_{-0.09}$ dex for the $\langle z \rangle = 2.26$ stack, and $8.35^{+0.14}_{-0.11}$ and $-2.72^{+0.08}_{-0.09}$ dex for the $\langle z \rangle = 3.15$ stack. Clearly, both direct values of $\mathrm{12+\log(O/H)}$ and $\mathrm{\log(Ar/O)}$ for both stacks are statistically consistent with their calibrated counterparts. This result demonstrates that the Ar/O strong-line calibration (Eq. \ref{eq:arCalib}) performs equally well on composite spectra as it does on the individual galaxies. We include these two direct abundance measurements on Figure \ref{fig:moneyplot}, and the direct-method oxygen abundances in Table~\ref{tab:table} as well.

\subsection{Composite Spectra}\label{subsection:composites}
Both composite spectra clearly exhibit subsolar Ar/O ratios, consistent with our expectation of delayed argon production by SNe Ia. This conclusion is legitimate despite the large errors of the Ar/O measurements introduced by the calibration, as the $\langle z \rangle = 2.26$ and $\langle z \rangle =3.15$  Ar/O ratios are both $3\sigma$ below the solar value. These representative samples indicate that galaxies during Cosmic Noon were young enough (i.e., having short enough star formation histories) that Ar/O is deficient relative to environments like the solar neighborhood, which have more extended star formation histories.

Our measurements are consistent with numerous other observations that suggest short star formation timescales present during Cosmic Noon. The observed relationship between oxygen and argon abundances obeys a similar relationship to that of the nebular oxygen to stellar iron ratio \citep[e.g.][]{Steidel2016, Topping2020, Cullen2021, Stanton2024}, which suggests $\alpha-$enhanced massive stars in $z\sim 2-3$ stellar populations. The star formation in these galaxies must have occurred on a short timescale to account for the excess $\alpha-$elements observed \citep{Strom2017}.
Emission line diagrams sensitive to ionization also support this physical picture; an offset toward higher $\mathrm{[OIII]\lambda5008/H\beta}$ at fixed $\mathrm{[NII]\lambda6858/H\alpha}$, or vice versa \citep[e.g.,][]{Shapley2019, Clarke2023, Shapley2025}, suggests the presence of a harder ionizing spectrum at fixed nebular oxygen abundance relative to what is observed in the local universe.
Such conditions occur at fixed $\alpha$-element abundance due to Fe deficiency (i.e. higher $\alpha/\mathrm{Fe}$), which generates a harder ionizing spectrum as a result of decreased UV opacity \citep{Steidel2016}. The Ar/O measurements presented here are another such abundance ratio that supports the conclusion of chemical youth, relative to the solar neighborhood. 

Quantitatively, the Ar/O ratios of our centroids are negatively offset from the results of \citet{Stanton2025} by $0.19^{+0.13}_{-0.12}$ and $0.19^{+0.12}_{-0.12}$ for the $\langle z \rangle = 2.26$ and $\langle z \rangle = 3.15$ stacks respectively. As indicated by the confidence interval, these differences are not statistically significant. Any offset can be attributed to the fact that the centroid from \citet{Stanton2025} is the inverse-variance weighted average of their individual detections, which would be less sensitive to weaker $[\mathrm{ArIII}]\lambda7137$ line strengths than our median stacks. It also may be a consequence of different ICFs or atomic datasets used between the calibrations of \citet{Sanders2025} and the direct calculations of \citet{Stanton2025}. 

\subsection{Individual Galaxies}
Of the 29 individual galaxies with detected $[\mathrm{ArIII}]\lambda7137$ lines, most have oxygen abundances as that are consistent with the composite spectra, indicating their representativeness of the sample as a whole. However, the centroids of the stacks appear to be systematically lower than the average of the individual galaxies in terms of their Ar/O abundance ratios. Quantitatively, the median Ar/O values of the individual galaxies with detections of $[\mathrm{ArIII}]\lambda7137$ were $\langle \mathrm{log(Ar/O)}\rangle = -2.66^{+0.03}_{-0.03}$ for the $\langle z \rangle = 2.26$ sample and $\langle \mathrm{log(Ar/O)}\rangle = -2.59^{+0.03}_{-0.04}$ for the $\langle z \rangle = 3.15$ sample, higher than the composite centroids by 0.03 and 0.1 dex, respectively. While this offset could be attributed to many causes, most likely it is simply a result of the high intrinsic scatter of the calibration. Additional biases introduced by the stacking process, including the non-additive nature of median statistics, would also bias the stacks toward a lower Ar/O ratio. Ultimately, given that our stack results are $1-3\sigma$ from the medians of the individual detections, our results are still valid for describing the typical degree of $\alpha-$enhancement of our samples. 

Another possible explanation is that the $[\mathrm{ArIII}]\lambda7137$ line is too weak to be detected at $3\sigma$ for galaxies with small Ar/O. These argon-deficient galaxies, while ignored in individual detections, would still contribute to the composite spectra, which can reproduce the weak $[\mathrm{ArIII}]\lambda7137$ line at higher S/N. However, we would expect the youngest galaxies with the lowest Ar/O ratios to have higher specific SFRs (i.e., SFR per unit stellar mass), and thus stronger lines overall. 

Our individual detections produce similar results to \citet{Rogers2025}, who report an average abundance ratio of $\log(\mathrm{Ar/O}) = -2.64\pm0.03$ dex from a sample of 19 individual galaxies. While the CECILIA survey samples from 36 total galaxies, only 19 of these have reported $\log(\mathrm{Ar/O})$ and $12+\log(\mathrm{O/H})$ values. We use this set of 19 for uniformity with our individual galaxies, in which we require both of these values to be defined. For reference, the inverse-variance weighted mean metallicity of these CECILIA galaxies is $12+\log(\mathrm{O/H}) = 8.27\pm0.04$ dex. Note that these values are drawn from the CECILIA dataset exclusively, and that the combined CECILIA and literature sample from \citet{Rogers2025} is significantly higher, at $\log(\mathrm{Ar/O}) = -2.56\pm0.04$ dex. The centroids of the individual AURORA detections for each bin are offset by a statistically insignificant value from the \citet{Rogers2025} CECILIA sample, and we ultimately come to similar conclusions regarding stellar nucleosynthetic pathways for argon production. 

\subsection{Comparison with GCE Modeling}\label{subsection:GCE}
To place our results into context with the MW, we compare our results with evolutionary models from \citet{Kobayashi2000, Kobayashi2020}. Since metal enrichment of the ISM is dependent on supernovae and matter inflow and outflow timescales, these models are constructed as functions of time in Gyr \citep{Kobayashi2020}. It is important to note that these models are designed within the context of MW multi-element abundance patterns localized within particular regions of the MW. Each model is given as a function of time across the same grid, meaning individual metal abundances and abundance ratios may be compared directly using these models by associating each value at each timestep with each other \citep{Stanton2025}. We include three models in Fig \ref{fig:moneyplot} of MW regions subject to different star-formation histories. These are the Solar Neighborhood, the Galactic Bulge with outflow, and the Galactic Bulge with the onset of galactic winds at $t = 3 \: \mathrm{Gyr}$. While the data from \citet{Stanton2025} is more consistent with the Solar Neighborhood model, we find our data more closely follows the GCE curves associated with the Galactic Bulge. 

We can determine the ages required to produce our observed ratios, taking advantage of the temporal dependence of the models from \citet{Kobayashi2020}.  Rather than compare the models to our ratios separately as a function of time, we take advantage of their shared time axis \citep[by construction, see][]{Kobayashi2020} to model $\log(\mathrm{Ar/O})$ as a function of $12 + \log(\mathrm{O/H})$, as was similarly done in \citet{Stanton2025}. For each model, we performed a 2D $\chi^2-$minimization for each pair of $\mathrm{\log(Ar/O)}$ and $\mathrm{12+\log(O/H)}$ points associated with each stack against the values predicted by the \citet{Kobayashi2020} models. The covariance of $\mathrm{\log(Ar/O)}$ and $\mathrm{12+\log(O/H)}$ was estimated using the Monte Carlo realizations generated previously to determine the confidence intervals of the individual ratios. Alongside the Bulge and Solar Neighborhood models shown in Figure \ref{fig:moneyplot}, we perform this analysis using the thick disk and halo models from \citet{Kobayashi2020} for completeness. 

Of these five models, we find the formation history that provides the best fit to our $\langle z \rangle = 2.26$ and $\langle z \rangle = 3.15$ composite measurements is associated with the Galactic Bulge with outflow model. Referencing the Bulge with outflow model, we find best-fit ages for the $\langle z \rangle = 2.26$ and $\langle z \rangle = 3.15$ stacks of $80^{+13}_{-2}$ Myr and $70^{+9}_{-2}$ Myr, respectively,  which are just shy of the model star formation timescales given \citep[$\tau_s = 0.1$Gyr,][]{Kobayashi2020}.
The Galactic Bulge model with the onset of galactic winds at $t = 3 \: \mathrm{Gyr}$ provides the next best fit to our Ar/O ratios. In the context of this other Bulge model, we infer ages, respectively, of $150^{+10}_{-16}$Myr and $120^{+16}_{-7}$ Myr for the $\langle z \rangle = 2.26$ and $\langle z \rangle = 3.15$ stacks. The short timescales with these Bulge models are consistent with the rapid formation of massive stars, aligning well with complementary $\alpha-$enhanced abundance ratios discussed above. These derived timescales are comparable to what has been found for star-forming $z\sim 2$ galaxies in a similar stellar mass range \citep[i.e., $>100$~Myr;][]{Shapley2005, Erb2006, Reddy2008}. 

The comparison of our data with these models comes with some important caveats. Firstly, we note that the average ages of the galaxies in the stacks, as inferred by their individual SED fits, are $388^{+102}_{-80}$ Myr for the $\langle z \rangle = 2.26$ stack and $401^{+114}_{-70}$ Myr for the $\langle z \rangle = 3.15$ stack. Since the \citet{Kobayashi2020} models are based on particular portions of the Milky Way (rather than the galaxy as a whole), and as such their assumed star formation histories may differ significantly from our chosen star formation history encoded in the SED fitting. In particular, the SED fitting assumes the delayed-$\tau$ SFR (see Section \ref{section:2}), which prescribes a smooth SFH throughout a galaxy's lifetime. However, many analyses of distant galactic populations suggest instead that more stochastic or ``bursty" star formation histories are necessary to explain scatter about the Star-Forming Main Sequence and the excess of galaxies with high UV luminosities at $z\sim10$ \citep{Clarke2024b, Clarke2025}. An early starburst, consistent with the formation of a galactic bulge, may explain the significantly younger ages estimated by the abundance patterns. We emphasize that neither ``age" is truly definitive, and both depend dramatically on the assumptions underlying each calculation. Firstly, the two ages are based on different observables, which trace different parts of the stellar population. Namely, the best-fit SED age is derived using the total integrated continuum starlight collected from the galaxy \citep{Kriek2009}, whereas the chemically-motivated ages are based on ISM abundance patterns, and thus are more sensitive to temperatures, emission-line atomic data, and ionization structure assumptions (which have their own issues, as discussed in Section~\ref{section:3}). This difference in estimated ages ultimately does not change the interpretation of our results, but may contribute to the offset between the models of \citep{Kobayashi2020} and our data.

Furthermore, the empirical calibrations used \citep{Sanders2025} do not account for depletion of oxygen into dust grains, which has been observed in star-forming regions in the local universe \citep{Esteban1998, Peimbert2010, PenaGuerrero2012, Amayo2021}. This depletion depends weakly on metallicity, and is expected to offset emission-line based oxygen abundances by 0.12 dex at Solar metallicity \citep{Amayo2021,Peimbert2010}. Accounting for dust depletion is relevant, as the GCE models of 
\citet{Kobayashi2020} predict elemental yields in the absence of dust.
Given that the confidence intervals of our composite measurements largely exceed this offset, and these composite spectra exhibit significantly subsolar metallicities, our conclusions are still valid despite not including oxygen dust depletion. As discussed in \citet{Stanton2025}, including contributions from depleted oxygen would offset all of our measurements toward lower Ar/O, thus further solidifying our claim of argon deficiency during Cosmic Noon compared to the local universe.
 
\section{Conclusions}

In summary, we have investigated the argon and oxygen emission line properties of 46 AURORA galaxies between $z = 2-2.6$ and $z = 3-3.5$. Through this analysis we come to the following conclusions:
\begin{itemize}
\item  Ar/O abundances may be estimated indirectly with a 2D function of log(Ar3O3) along with log(O32) as an ionization structure indicator, correcting for argon abundances not probed by the Ar3O3 ratio alone. This function works equally well on individual galaxies and composite spectra.
\item Our composite median-stacked spectra and our individual detections exhibit significantly deficient Ar/O ratios compared to the solar neighborhood, indicating the ratio's consistency with more typical identifiers of $\alpha-$enhancement.
\item Our representative samples align well with GCE modeling of the Galactic Bulge, consistent with a bulge-dominated star formation history present during Cosmic Noon. However, to obtain more robust GCE-based timescales, models more tuned to the chemical abundances of Cosmic Noon galaxies (rather than the MW) are needed to draw stronger conclusions on the role of galactic bulges in facilitating star formation.  
\end{itemize}

Our results for both composite spectra and individual galaxies are subsolar, consistent with the delayed production of additional argon by SNe Ia to supplement that produced by CCSNe. Argon thus may be used in tandem with available iron abundances to constrain GCE modeling and SNe Ia timescales, providing a more complete picture of the role of SNe Ia in the metal enrichment of the ISM across cosmic time. Direct measurements of Ar/O abundances will help tighten the empirical Ar/O calibration of Eq. \ref{eq:arCalib} for the $z\sim 2-4$ redshift range, improving the quality of its estimations. Comparisons of the Ar/O abundance ratio with other identifiers of star formation, such as the gas fraction or the specific star formation rate (sSFR), will help elucidate further the relative contributions of CCSNe and SNe Ia toward Ar/O. Further investigations of larger samples that cover greater redshift ranges will be needed to resolve the time-evolution of the Ar/O abundance ratio during Cosmic Noon itself. Such analyses are instrumental for fully exploring the uses of argon in understanding the star-forming conditions of the early Universe in future spectroscopic studies.

\section*{Acknowledgements}
This work is based on observations made with the NASA/ESA/CSA James Webb Space Telescope. The data were
obtained from the Mikulski Archive for Space Telescopes at
the Space Telescope Science Institute, which is operated by the
Association of Universities for Research in Astronomy, Inc.,
under NASA contract NAS5-03127 for JWST.  The specific observations analyzed can be accessed via \dataset[DOI: 10.17909/6mza-5q55]{https://archive.stsci.edu/doi/resolve/resolve.html?doi=10.17909/6mza-5q55}.
We also acknowledge support from NASA grant JWST-GO-01914. TMS and FC acknowledges support from a UKRI Frontier Research Guarantee Grant (PI Cullen; grant reference: EP/X021025/1). TJ gratefully acknowledges support from NASA under grant 80NSSC23K1132, and from a UC Davis Chancellor's Fellowship.

\clearpage
\bibliographystyle{aasjournal}
\bibliography{refs.bib}

\begin{thebibliography}{}
\expandafter\ifx\csname natexlab\endcsname\relax\def\natexlab#1{#1}\fi
\providecommand{\url}[1]{\href{#1}{#1}}
\providecommand{\dodoi}[1]{doi:~\href{http://doi.org/#1}{\nolinkurl{#1}}}
\providecommand{\doeprint}[1]{\href{http://ascl.net/#1}{\nolinkurl{http://ascl.net/#1}}}
\providecommand{\doarXiv}[1]{\href{https://arxiv.org/abs/#1}{\nolinkurl{https://arxiv.org/abs/#1}}}

\bibitem[{{Amayo} {et~al.}(2021){Amayo}, {Delgado-Inglada}, \& {Stasi{\'n}ska}}]{Amayo2021}
{Amayo}, A., {Delgado-Inglada}, G., \& {Stasi{\'n}ska}, G. 2021, \mnras, 505, 2361, \dodoi{10.1093/mnras/stab1467}

\bibitem[{{Arellano-C{\'o}rdova} \& {Rodr{\'\i}guez}(2020)}]{ArellanoCordova2020}
{Arellano-C{\'o}rdova}, K.~Z., \& {Rodr{\'\i}guez}, M. 2020, \mnras, 497, 672, \dodoi{10.1093/mnras/staa1759}

\bibitem[{{Arellano-C{\'o}rdova} {et~al.}(2024){Arellano-C{\'o}rdova}, {Berg}, {Mingozzi}, {James}, {Rogers}, {Skillman}, {Cullen}, {Alexander}, {Amor{\'\i}n}, {Chisholm}, {Hayes}, {Heckman}, {Hernandez}, {Kumari}, {Leitherer}, {Martin}, {Maseda}, {Nanayakkara}, {Parker}, {Ravindranath}, {Strom}, {Vincenzo}, \& {Wofford}}]{Arellano-Cordova2024}
{Arellano-C{\'o}rdova}, K.~Z., {Berg}, D.~A., {Mingozzi}, M., {et~al.} 2024, \apj, 968, 98, \dodoi{10.3847/1538-4357/ad34cf}

\bibitem[{{Arnaboldi} {et~al.}(2022){Arnaboldi}, {Bhattacharya, Souradeep}, {Gerhard, Ortwin}, {Kobayashi, Chiaki}, {Freeman, Kenneth C.}, {Caldwell, Nelson}, {Hartke, Johanna}, {McConnachie, Alan}, \& {Guhathakurta, Puragra}}]{Arnaboldi2022}
{Arnaboldi}, M., {Bhattacharya, Souradeep}, {Gerhard, Ortwin}, {et~al.} 2022, A\&A, 666, A109, \dodoi{10.1051/0004-6361/202244258}

\bibitem[{Asano {et~al.}(2013)Asano, Takeuchi, Hirashita, \& Inoue}]{Asano2013}
Asano, R.~S., Takeuchi, T.~T., Hirashita, H., \& Inoue, A.~K. 2013, Earth, Planets and Space, 65, 213, \dodoi{10.5047/eps.2012.04.014}

\bibitem[{{Asplund} {et~al.}(2021){Asplund}, {Amarsi}, \& {Grevesse}}]{Asplund2021}
{Asplund}, M., {Amarsi}, A.~M., \& {Grevesse}, N. 2021, \aap, 653, A141, \dodoi{10.1051/0004-6361/202140445}

\bibitem[{{Baldwin} {et~al.}(1981){Baldwin}, {Phillips}, \& {Terlevich}}]{BPT}
{Baldwin}, J.~A., {Phillips}, M.~M., \& {Terlevich}, R. 1981, \pasp, 93, 5, \dodoi{10.1086/130766}

\bibitem[{{Benton} {et~al.}(2024){Benton}, {Nelson}, {Miller}, {Bezanson}, {Gibson}, {Hartley}, {Martorano}, {Price}, {Suess}, {van der Wel}, {van Dokkum}, {Weaver}, \& {Whitaker}}]{Benton2024}
{Benton}, C.~E., {Nelson}, E.~J., {Miller}, T.~B., {et~al.} 2024, \apjl, 974, L28, \dodoi{10.3847/2041-8213/ad7e27}

\bibitem[{{Berg} {et~al.}(2021){Berg}, {Chisholm}, {Erb}, {Skillman}, {Pogge}, \& {Olivier}}]{Berg2021}
{Berg}, D.~A., {Chisholm}, J., {Erb}, D.~K., {et~al.} 2021, \apj, 922, 170, \dodoi{10.3847/1538-4357/ac141b}

\bibitem[{{Berg} {et~al.}(2025){Berg}, {Naidu}, {Chisholm}, {Atek}, {Fujimoto}, {Kokorev}, {Furtak}, {Kobayashi}, {Schaerer}, {Adamo}, {Fei}, {Korber}, {Matthee}, {Marques-Chaves}, {Martinez}, {Mcquinn}, {Mu{\~n}oz}, {Oesch}, {Stark}, {Stephenson}, \& {Hsiao}}]{Berg2025}
{Berg}, D.~A., {Naidu}, R.~P., {Chisholm}, J., {et~al.} 2025, arXiv e-prints, arXiv:2511.13591, \dodoi{10.48550/arXiv.2511.13591}

\bibitem[{Beverage {et~al.}(2023)Beverage, Kriek, Conroy, Sandford, Bezanson, Franx, van~der Wel, \& Weisz}]{Beverage2023}
Beverage, A.~G., Kriek, M., Conroy, C., {et~al.} 2023, The Astrophysical Journal, 948, 140, \dodoi{10.3847/1538-4357/acc176}

\bibitem[{{Beverage} {et~al.}(2025){Beverage}, {Slob}, {Kriek}, {Conroy}, {Barro}, {Bezanson}, {Brammer}, {Cheng}, {de Graaff}, {F{\"o}rster Schreiber}, {Franx}, {Lorenz}, {Mancera Pi{\~n}a}, {Marchesini}, {Muzzin}, {Newman}, {Price}, {Shapley}, {Stefanon}, {Suess}, {van Dokkum}, {Weinberg}, \& {Weisz}}]{Beverage2025}
{Beverage}, A.~G., {Slob}, M., {Kriek}, M., {et~al.} 2025, \apj, 979, 249, \dodoi{10.3847/1538-4357/ad96b6}

\bibitem[{{Bhattacharya} {et~al.}(2025){Bhattacharya}, {Arnaboldi}, {Gerhard}, {Kobayashi}, \& {Saha}}]{Bhattacharya2025}
{Bhattacharya}, S., {Arnaboldi}, M., {Gerhard}, O., {Kobayashi}, C., \& {Saha}, K. 2025, \apjl, 983, L30, \dodoi{10.3847/2041-8213/adc735}

\bibitem[{Bhattacharya {et~al.}(2025)Bhattacharya, Arnaboldi, Kobayashi, Gerhard, \& Saha}]{Bhattacharya2025b}
Bhattacharya, S., Arnaboldi, M., Kobayashi, C., Gerhard, O., \& Saha, K. 2025, Mass-dependent chemical enrichment sequences of SDSS star-forming galaxies out to $z\sim 0.3$ revealed by direct O \& Ar abundances.
\newblock \doarXiv{2505.01896}

\bibitem[{Calzetti {et~al.}(2000)Calzetti, Armus, Bohlin, Kinney, Koornneef, \& Storchi-Bergmann}]{Calzetti2000}
Calzetti, D., Armus, L., Bohlin, R.~C., {et~al.} 2000, The Astrophysical Journal, 533, 682, \dodoi{10.1086/308692}

\bibitem[{{Campbell} {et~al.}(1986){Campbell}, {Terlevich}, \& {Melnick}}]{Campbell1986}
{Campbell}, A., {Terlevich}, R., \& {Melnick}, J. 1986, \mnras, 223, 811, \dodoi{10.1093/mnras/223.4.811}

\bibitem[{{Cardelli} {et~al.}(1989){Cardelli}, {Clayton}, \& {Mathis}}]{Cardelli1989}
{Cardelli}, J.~A., {Clayton}, G.~C., \& {Mathis}, J.~S. 1989, \apj, 345, 245, \dodoi{10.1086/167900}

\bibitem[{{Carnall} {et~al.}(2024){Carnall}, {Cullen}, {McLure}, {McLeod}, {Begley}, {Donnan}, {Dunlop}, {Shapley}, {Rowlands}, {Almaini}, {Arellano-C{\'o}rdova}, {Barrufet}, {Cimatti}, {Ellis}, {Grogin}, {Hamadouche}, {Illingworth}, {Koekemoer}, {Leung}, {Lovell}, {P{\'e}rez-Gonz{\'a}lez}, {Santini}, {Stanton}, \& {Wild}}]{EXCELS}
{Carnall}, A.~C., {Cullen}, F., {McLure}, R.~J., {et~al.} 2024, \mnras, 534, 325, \dodoi{10.1093/mnras/stae2092}

\bibitem[{Chabrier(2003)}]{Chabrier2003}
Chabrier, G. 2003, Publications of the Astronomical Society of the Pacific, 115, 763, \dodoi{10.1086/376392}

\bibitem[{{Clarke} {et~al.}(2026){Clarke}, {Lam}, {Shapley}, {Topping}, {Brammer}, {Sanders}, {Reddy}, \& {Karthikeyan}}]{Clarke2026}
{Clarke}, L., {Lam}, N., {Shapley}, A.~E., {et~al.} 2026, arXiv e-prints, arXiv:2601.11092, \dodoi{10.48550/arXiv.2601.11092}

\bibitem[{{Clarke} {et~al.}(2025){Clarke}, {Shapley}, {Lam}, {Topping}, {Brammer}, {Sanders}, {Reddy}, \& {Karthikeyan}}]{Clarke2025}
{Clarke}, L., {Shapley}, A.~E., {Lam}, N., {et~al.} 2025, arXiv e-prints, arXiv:2510.06681, \dodoi{10.48550/arXiv.2510.06681}

\bibitem[{{Clarke} {et~al.}(2024){Clarke}, {Shapley}, {Sanders}, {Topping}, {Brammer}, {Bento}, {Reddy}, \& {Kehoe}}]{Clarke2024b}
{Clarke}, L., {Shapley}, A.~E., {Sanders}, R.~L., {et~al.} 2024, \apj, 977, 133, \dodoi{10.3847/1538-4357/ad8ba4}

\bibitem[{Clarke {et~al.}(2023)Clarke, Shapley, Sanders, Topping, Jones, Kriek, Reddy, Stark, \& Tang}]{Clarke2023}
Clarke, L., Shapley, A., Sanders, R.~L., {et~al.} 2023, The Astrophysical Journal, 957, 81, \dodoi{10.3847/1538-4357/acfedb}

\bibitem[{{Clarke} {et~al.}(2023){Clarke}, {Shapley}, {Sanders}, {Topping}, {Jones}, {Kriek}, {Reddy}, {Stark}, \& {Tang}}]{Clarke2024}
{Clarke}, L., {Shapley}, A., {Sanders}, R.~L., {et~al.} 2023, \apj, 957, 81, \dodoi{10.3847/1538-4357/acfedb}

\bibitem[{Conroy {et~al.}(2009)Conroy, Gunn, \& White}]{Conroy2009}
Conroy, C., Gunn, J.~E., \& White, M. 2009, The Astrophysical Journal, 699, 486, \dodoi{10.1088/0004-637X/699/1/486}

\bibitem[{Conroy \& van Dokkum(2012)}]{Conroy2012}
Conroy, C., \& van Dokkum, P.~G. 2012, The Astrophysical Journal, 760, 71, \dodoi{10.1088/0004-637X/760/1/71}

\bibitem[{{Cullen} {et~al.}(2021){Cullen}, {Shapley}, {McLure}, {Dunlop}, {Sanders}, {Topping}, {Reddy}, {Amor{\'\i}n}, {Begley}, {Bolzonella}, {Calabr{\`o}}, {Carnall}, {Castellano}, {Cimatti}, {Cirasuolo}, {Cresci}, {Fontana}, {Fontanot}, {Garilli}, {Guaita}, {Hamadouche}, {Hathi}, {Mannucci}, {McLeod}, {Pentericci}, {Saxena}, {Talia}, \& {Zamorani}}]{Cullen2021}
{Cullen}, F., {Shapley}, A.~E., {McLure}, R.~J., {et~al.} 2021, \mnras, 505, 903, \dodoi{10.1093/mnras/stab1340}

\bibitem[{{Curti} {et~al.}(2017){Curti}, {Cresci}, {Mannucci}, {Marconi}, {Maiolino}, \& {Esposito}}]{Curti2017}
{Curti}, M., {Cresci}, G., {Mannucci}, F., {et~al.} 2017, \mnras, 465, 1384, \dodoi{10.1093/mnras/stw2766}

\bibitem[{{Curti} {et~al.}(2025){Curti}, {Cataldi}, {Belfiore}, {Moreschini}, {Arnaboldi}, {Chru{\'s}li{\'n}ska}, {Mannucci}, {Marconi}, {D'Amato}, {Carniani}, {Baker}, {De Cia}, {Kumari}, {Amiri}, {Cresci}, {Kobayashi}, {Cullen}, {Feltre}, \& {Maiolino}}]{Curti2025}
{Curti}, M., {Cataldi}, E., {Belfiore}, F., {et~al.} 2025, arXiv e-prints, arXiv:2509.06622, \dodoi{10.48550/arXiv.2509.06622}

\bibitem[{{Davies} {et~al.}(2021){Davies}, {F{\"o}rster Schreiber}, {Genzel}, {Shimizu}, {Davies}, {Schruba}, {Tacconi}, {{\"U}bler}, {Wisnioski}, {Wuyts}, {Fossati}, {Herrera-Camus}, {Lutz}, {Mendel}, {Naab}, {Price}, {Renzini}, {Wilman}, {Beifiori}, {Belli}, {Burkert}, {Chan}, {Contursi}, {Fabricius}, {Lee}, {Saglia}, \& {Sternberg}}]{Davies2021}
{Davies}, R.~L., {F{\"o}rster Schreiber}, N.~M., {Genzel}, R., {et~al.} 2021, \apj, 909, 78, \dodoi{10.3847/1538-4357/abd551}

\bibitem[{{de Graaff} {et~al.}(2024){de Graaff}, {Rix}, {Carniani}, {Suess}, {Charlot}, {Curtis-Lake}, {Arribas}, {Baker}, {Boyett}, {Bunker}, {Cameron}, {Chevallard}, {Curti}, {Eisenstein}, {Franx}, {Hainline}, {Hausen}, {Ji}, {Johnson}, {Jones}, {Maiolino}, {Maseda}, {Nelson}, {Parlanti}, {Rawle}, {Robertson}, {Tacchella}, {{\"U}bler}, {Williams}, {Willmer}, \& {Willott}}]{deGraaff2024}
{de Graaff}, A., {Rix}, H.-W., {Carniani}, S., {et~al.} 2024, \aap, 684, A87, \dodoi{10.1051/0004-6361/202347755}

\bibitem[{{Erb} {et~al.}(2006{\natexlab{a}}){Erb}, {Shapley}, {Pettini}, {Steidel}, {Reddy}, \& {Adelberger}}]{Erb2006b}
{Erb}, D.~K., {Shapley}, A.~E., {Pettini}, M., {et~al.} 2006{\natexlab{a}}, \apj, 644, 813, \dodoi{10.1086/503623}

\bibitem[{{Erb} {et~al.}(2006{\natexlab{b}}){Erb}, {Steidel}, {Shapley}, {Pettini}, {Reddy}, \& {Adelberger}}]{Erb2006}
{Erb}, D.~K., {Steidel}, C.~C., {Shapley}, A.~E., {et~al.} 2006{\natexlab{b}}, \apj, 646, 107, \dodoi{10.1086/504891}

\bibitem[{{Esteban} {et~al.}(1998){Esteban}, {Peimbert}, {Torres-Peimbert}, \& {Escalante}}]{Esteban1998}
{Esteban}, C., {Peimbert}, M., {Torres-Peimbert}, S., \& {Escalante}, V. 1998, \mnras, 295, 401, \dodoi{10.1046/j.1365-8711.1998.01335.x}

\bibitem[{{F{\"o}rster Schreiber} \& {Wuyts}(2020)}]{ForsterSchreiber2020}
{F{\"o}rster Schreiber}, N.~M., \& {Wuyts}, S. 2020, \araa, 58, 661, \dodoi{10.1146/annurev-astro-032620-021910}

\bibitem[{{Froese Fischer} \& {Tachiev}(2004)}]{Froese2004}
{Froese Fischer}, C., \& {Tachiev}, G. 2004, Atomic Data and Nuclear Data Tables, 87, 1, \dodoi{10.1016/j.adt.2004.02.001}

\bibitem[{Gordon {et~al.}(2003)Gordon, Clayton, Misselt, Landolt, \& Wolff}]{Gordon2003}
Gordon, K.~D., Clayton, G.~C., Misselt, K.~A., Landolt, A.~U., \& Wolff, M.~J. 2003, The Astrophysical Journal, 594, 279, \dodoi{10.1086/376774}

\bibitem[{{Grevesse} {et~al.}(2010){Grevesse}, {Asplund}, {Sauval}, \& {Scott}}]{Asplund2010}
{Grevesse}, N., {Asplund}, M., {Sauval}, A.~J., \& {Scott}, P. 2010, \apss, 328, 179, \dodoi{10.1007/s10509-010-0288-z}

\bibitem[{{Isobe} {et~al.}(2025){Isobe}, {Maiolino}, {Ji}, {D'Eugenio}, {Simmonds}, {Scholtz}, {Juod{\v{z}}balis}, {Saxena}, {Witstok}, {Kobayashi}, {Vanni}, {Salvadori}, {Watanabe}, {Monty}, {Belokurov}, {Feltre}, {McClymont}, {Tacchella}, {Curti}, {{\"U}bler}, {Charlot}, {Bunker}, {Chevallard}, {Curtis-Lake}, {Kumari}, {Rinaldi}, {Robertson}, {Williams}, \& {Willott}}]{Isobe2025}
{Isobe}, Y., {Maiolino}, R., {Ji}, X., {et~al.} 2025, arXiv e-prints, arXiv:2509.18055, \dodoi{10.48550/arXiv.2509.18055}

\bibitem[{{Izotov} {et~al.}(2006){Izotov}, {Stasi{\'n}ska}, {Meynet}, {Guseva}, \& {Thuan}}]{Izotov2006}
{Izotov}, Y.~I., {Stasi{\'n}ska}, G., {Meynet}, G., {Guseva}, N.~G., \& {Thuan}, T.~X. 2006, \aap, 448, 955, \dodoi{10.1051/0004-6361:20053763}

\bibitem[{{Jafariyazani} {et~al.}(2025){Jafariyazani}, {Newman}, {Mobasher}, {Belli}, {Ellis}, \& {Faisst}}]{Jafariyazani2025}
{Jafariyazani}, M., {Newman}, A.~B., {Mobasher}, B., {et~al.} 2025, \apj, 986, 148, \dodoi{10.3847/1538-4357/addbdc}

\bibitem[{Jones {et~al.}(2018)Jones, Stark, \& Ellis}]{Jones2018}
Jones, T., Stark, D.~P., \& Ellis, R.~S. 2018, The Astrophysical Journal, 863, 191, \dodoi{10.3847/1538-4357/aad37f}

\bibitem[{{Kashino} {et~al.}(2017){Kashino}, {Silverman}, {Sanders}, {Kartaltepe}, {Daddi}, {Renzini}, {Valentino}, {Rodighiero}, {Juneau}, {Kewley}, {Zahid}, {Arimoto}, {Nagao}, {Chu}, {Sugiyama}, {Civano}, {Ilbert}, {Kajisawa}, {Le F{\`e}vre}, {Maier}, {Masters}, {Miyaji}, {Onodera}, {Puglisi}, \& {Taniguchi}}]{Kashino2017}
{Kashino}, D., {Silverman}, J.~D., {Sanders}, D., {et~al.} 2017, \apj, 835, 88, \dodoi{10.3847/1538-4357/835/1/88}

\bibitem[{{Kehoe} {et~al.}(2025){Kehoe}, {Shapley}, {Sanders}, {Reddy}, {Topping}, {Lam}, {Clarke}, {Cullen}, {Ellis}, {Forster Schreiber}, {Jones}, {Khostovan}, {McLeod}, {McLure}, {Narayanan}, {Oesch}, \& {Pahl}}]{Kehoe2025}
{Kehoe}, E., {Shapley}, A.~E., {Sanders}, R.~L., {et~al.} 2025, arXiv e-prints, arXiv:2506.17381, \dodoi{10.48550/arXiv.2506.17381}

\bibitem[{{Khostovan} {et~al.}(2025){Khostovan}, {Sanders}, {Shapley}, {Topping}, {Reddy}, {Garcia}, {Berg}, {Clarke}, {Cullen}, {Ellis}, {F{\"o}rster Schreiber}, {Glazebrook}, {Jones}, {McLeod}, {Pahl}, {Pettini}, \& {Torrey}}]{Khostovan2025}
{Khostovan}, A.~A., {Sanders}, R.~L., {Shapley}, A.~E., {et~al.} 2025, arXiv e-prints, arXiv:2512.16989, \dodoi{10.48550/arXiv.2512.16989}

\bibitem[{{Kisielius} {et~al.}(2009){Kisielius}, {Storey}, {Ferland}, \& {Keenan}}]{Kisielius2009}
{Kisielius}, R., {Storey}, P.~J., {Ferland}, G.~J., \& {Keenan}, F.~P. 2009, \mnras, 397, 903, \dodoi{10.1111/j.1365-2966.2009.14989.x}

\bibitem[{{Kobayashi} {et~al.}(2023){Kobayashi}, {Bhattacharya}, {Arnaboldi}, \& {Gerhard}}]{Kobayashi2023}
{Kobayashi}, C., {Bhattacharya}, S., {Arnaboldi}, M., \& {Gerhard}, O. 2023, \apjl, 956, L14, \dodoi{10.3847/2041-8213/acf7c7}

\bibitem[{{Kobayashi} {et~al.}(2020){Kobayashi}, {Karakas}, \& {Lugaro}}]{Kobayashi2020}
{Kobayashi}, C., {Karakas}, A.~I., \& {Lugaro}, M. 2020, \apj, 900, 179, \dodoi{10.3847/1538-4357/abae65}

\bibitem[{Kobayashi {et~al.}(2020)Kobayashi, Leung, \& Nomoto}]{Kobayashi2020a}
Kobayashi, C., Leung, S.-C., \& Nomoto, K. 2020, The Astrophysical Journal, 895, 138, \dodoi{10.3847/1538-4357/ab8e44}

\bibitem[{{Kobayashi} {et~al.}(2000){Kobayashi}, {Tsujimoto}, \& {Nomoto}}]{Kobayashi2000}
{Kobayashi}, C., {Tsujimoto}, T., \& {Nomoto}, K. 2000, \apj, 539, 26, \dodoi{10.1086/309195}

\bibitem[{Kriek {et~al.}(2009)Kriek, van Dokkum, Labbé, Franx, Illingworth, Marchesini, \& Quadri}]{Kriek2009}
Kriek, M., van Dokkum, P.~G., Labbé, I., {et~al.} 2009, The Astrophysical Journal, 700, 221, \dodoi{10.1088/0004-637X/700/1/221}

\bibitem[{{Kriek} {et~al.}(2019){Kriek}, {Price}, {Conroy}, {Suess}, {Mowla}, {Pasha}, {Bezanson}, {van Dokkum}, \& {Barro}}]{Kriek2019}
{Kriek}, M., {Price}, S.~H., {Conroy}, C., {et~al.} 2019, \apjl, 880, L31, \dodoi{10.3847/2041-8213/ab2e75}

\bibitem[{{Lam} {et~al.}(2025){Lam}, {Shapley}, {Sanders}, {Do}, {Jones}, {Coil}, {Kriek}, {Mobasher}, {Reddy}, {Siana}, \& {Clarke}}]{Lam2025}
{Lam}, N., {Shapley}, A.~E., {Sanders}, R.~L., {et~al.} 2025, arXiv e-prints, arXiv:2506.22547, \dodoi{10.48550/arXiv.2506.22547}

\bibitem[{{Lodders} {et~al.}(2025){Lodders}, {Bergemann}, \& {Palme}}]{Lodders2025}
{Lodders}, K., {Bergemann}, M., \& {Palme}, H. 2025, \ssr, 221, 23, \dodoi{10.1007/s11214-025-01146-w}

\bibitem[{{Luridiana} {et~al.}(2015){Luridiana}, {Morisset}, \& {Shaw}}]{Luridiana2015}
{Luridiana}, V., {Morisset}, C., \& {Shaw}, R.~A. 2015, \aap, 573, A42, \dodoi{10.1051/0004-6361/201323152}

\bibitem[{{Madau} \& {Dickinson}(2014)}]{Piero2014}
{Madau}, P., \& {Dickinson}, M. 2014, \araa, 52, 415, \dodoi{10.1146/annurev-astro-081811-125615}

\bibitem[{{Maiolino} {et~al.}(2008){Maiolino}, {Nagao}, {Grazian}, {Cocchia}, {Marconi}, {Mannucci}, {Cimatti}, {Pipino}, {Ballero}, {Calura}, {Chiappini}, {Fontana}, {Granato}, {Matteucci}, {Pastorini}, {Pentericci}, {Risaliti}, {Salvati}, \& {Silva}}]{Maiolino2008}
{Maiolino}, R., {Nagao}, T., {Grazian}, A., {et~al.} 2008, \aap, 488, 463, \dodoi{10.1051/0004-6361:200809678}

\bibitem[{{Mannucci} {et~al.}(2006){Mannucci}, {Della Valle}, \& {Panagia}}]{Mannucci2006}
{Mannucci}, F., {Della Valle}, M., \& {Panagia}, N. 2006, \mnras, 370, 773, \dodoi{10.1111/j.1365-2966.2006.10501.x}

\bibitem[{{Maoz} {et~al.}(2012){Maoz}, {Mannucci}, \& {Brandt}}]{Maoz2012}
{Maoz}, D., {Mannucci}, F., \& {Brandt}, T.~D. 2012, \mnras, 426, 3282, \dodoi{10.1111/j.1365-2966.2012.21871.x}

\bibitem[{{Martinez} {et~al.}(2025){Martinez}, {Berg}, {James}, {Arellano-C{\'o}rdova}, {Stark}, {Senchyna}, {Skillman}, {Rogers}, \& {Chisholm}}]{Martinez2025}
{Martinez}, Z., {Berg}, D.~A., {James}, B.~L., {et~al.} 2025, \apj, 995, 204, \dodoi{10.3847/1538-4357/ae17c6}

\bibitem[{{Maseda} {et~al.}(2017){Maseda}, {Brinchmann}, {Franx}, {Bacon}, {Bouwens}, {Schmidt}, {Boogaard}, {Contini}, {Feltre}, {Inami}, {Kollatschny}, {Marino}, {Richard}, {Verhamme}, \& {Wisotzki}}]{Maseda2017}
{Maseda}, M.~V., {Brinchmann}, J., {Franx}, M., {et~al.} 2017, \aap, 608, A4, \dodoi{10.1051/0004-6361/201730985}

\bibitem[{{Matteucci} \& {Greggio}(1986)}]{Matteuci1986}
{Matteucci}, F., \& {Greggio}, L. 1986, \aap, 154, 279

\bibitem[{{M{\'e}ndez-Delgado} {et~al.}(2024){M{\'e}ndez-Delgado}, {Kreckel}, {Esteban}, {Garc{\'\i}a-Rojas}, {Carigi}, {Sander}, {Palla}, {Chru{\'s}li{\'n}ska}, {De Looze}, {Rela{\~n}o}, {van der Giessen}, {Reyes-Rodr{\'\i}guez}, \& {S{\'a}nchez}}]{MendezDelgado2024}
{M{\'e}ndez-Delgado}, J.~E., {Kreckel}, K., {Esteban}, C., {et~al.} 2024, \aap, 690, A248, \dodoi{10.1051/0004-6361/202450928}

\bibitem[{{Monty} {et~al.}(2025){Monty}, {Strom}, {Stanton}, {Chru{\'s}li{\'n}ska}, {Cullen}, {Kobayashi}, {Starkenburg}, {Bhattacharya}, {Sanders}, \& {Gieles}}]{Monty2025}
{Monty}, S., {Strom}, A.~L., {Stanton}, T.~M., {et~al.} 2025, \mnras, \dodoi{10.1093/mnras/staf1213}

\bibitem[{{Munoz Burgos} {et~al.}(2009){Munoz Burgos}, {Loch}, {Ballance}, \& {Boivin}}]{Munoz2009}
{Munoz Burgos}, J.~M., {Loch}, S.~D., {Ballance}, C.~P., \& {Boivin}, R.~F. 2009, \aap, 500, 1253, \dodoi{10.1051/0004-6361/200911743}

\bibitem[{{Nomoto} {et~al.}(2013){Nomoto}, {Kobayashi}, \& {Tominaga}}]{Nomoto2013}
{Nomoto}, K., {Kobayashi}, C., \& {Tominaga}, N. 2013, \araa, 51, 457, \dodoi{10.1146/annurev-astro-082812-140956}

\bibitem[{{Osterbrock} {et~al.}(1992){Osterbrock}, {Tran}, \& {Veilleux}}]{Osterbrock1992}
{Osterbrock}, D.~E., {Tran}, H.~D., \& {Veilleux}, S. 1992, \apj, 389, 305, \dodoi{10.1086/171206}

\bibitem[{Papovich {et~al.}(2022)Papovich, Simons, Estrada-Carpenter, Matharu, Momcheva, Trump, Backhaus, Brammer, Cleri, Finkelstein, Giavalisco, Ji, Jung, Kewley, Nicholls, Pirzkal, Rafelski, \& Weiner}]{Papovich2022}
Papovich, C., Simons, R.~C., Estrada-Carpenter, V., {et~al.} 2022, The Astrophysical Journal, 937, 22, \dodoi{10.3847/1538-4357/ac8058}

\bibitem[{{Pe{\~n}a-Guerrero} {et~al.}(2012){Pe{\~n}a-Guerrero}, {Peimbert}, \& {Peimbert}}]{PenaGuerrero2012}
{Pe{\~n}a-Guerrero}, M.~A., {Peimbert}, A., \& {Peimbert}, M. 2012, \apjl, 756, L14, \dodoi{10.1088/2041-8205/756/1/L14}

\bibitem[{{Peimbert} \& {Peimbert}(2010)}]{Peimbert2010}
{Peimbert}, A., \& {Peimbert}, M. 2010, \apj, 724, 791, \dodoi{10.1088/0004-637X/724/1/791}

\bibitem[{{P{\'e}rez-Montero} \& {D{\'\i}az}(2005)}]{PerezMontero2005}
{P{\'e}rez-Montero}, E., \& {D{\'\i}az}, A.~I. 2005, \mnras, 361, 1063, \dodoi{10.1111/j.1365-2966.2005.09263.x}

\bibitem[{Pettini {et~al.}(2002)Pettini, Rix, Steidel, Adelberger, Hunt, \& Shapley}]{Pettini2002}
Pettini, M., Rix, S.~A., Steidel, C.~C., {et~al.} 2002, The Astrophysical Journal, 569, 742, \dodoi{10.1086/339355}

\bibitem[{{Pilyugin} \& {Grebel}(2016)}]{Pilyugin2016}
{Pilyugin}, L.~S., \& {Grebel}, E.~K. 2016, \mnras, 457, 3678, \dodoi{10.1093/mnras/stw238}

\bibitem[{{Reddy} {et~al.}(2008){Reddy}, {Steidel}, {Pettini}, {Adelberger}, {Shapley}, {Erb}, \& {Dickinson}}]{Reddy2008}
{Reddy}, N.~A., {Steidel}, C.~C., {Pettini}, M., {et~al.} 2008, \apjs, 175, 48, \dodoi{10.1086/521105}

\bibitem[{Reddy {et~al.}(2023)Reddy, Topping, Sanders, Shapley, \& Brammer}]{Reddy2023}
Reddy, N.~A., Topping, M.~W., Sanders, R.~L., Shapley, A.~E., \& Brammer, G. 2023, The Astrophysical Journal, 948, 83, \dodoi{10.3847/1538-4357/acc869}

\bibitem[{Reddy {et~al.}(2018)Reddy, Oesch, Bouwens, Montes, Illingworth, Steidel, van Dokkum, Atek, Carollo, Cibinel, Holden, Labbé, Magee, Morselli, Nelson, \& Wilkins}]{Reddy2018}
Reddy, N.~A., Oesch, P.~A., Bouwens, R.~J., {et~al.} 2018, The Astrophysical Journal, 853, 56, \dodoi{10.3847/1538-4357/aaa3e7}

\bibitem[{{Reddy} {et~al.}(2023){Reddy}, {Sanders}, {Shapley}, {Topping}, {Kriek}, {Coil}, {Mobasher}, {Siana}, \& {Rezaee}}]{Reddy2023b}
{Reddy}, N.~A., {Sanders}, R.~L., {Shapley}, A.~E., {et~al.} 2023, \apj, 951, 56, \dodoi{10.3847/1538-4357/acd0b1}

\bibitem[{{Reddy} {et~al.}(2026){Reddy}, {Shapley}, {Sanders}, {Topping}, {Ellis}, {Pettini}, {Brammer}, {Cullen}, {F{\"o}rster Schreiber}, {Khostovan}, {McLeod}, {McLure}, {Narayanan}, {Oesch}, {Pahl}, {Steidel}, \& {Berg}}]{Reddy2025}
{Reddy}, N.~A., {Shapley}, A.~E., {Sanders}, R.~L., {et~al.} 2026, \apj, 999, 15, \dodoi{10.3847/1538-4357/ae38da}

\bibitem[{Rodríguez \& Rubin(2005)}]{Rodriguez2005}
Rodríguez, M., \& Rubin, R.~H. 2005, The Astrophysical Journal, 626, 900, \dodoi{10.1086/429958}

\bibitem[{{Rogers} {et~al.}(2022){Rogers}, {Skillman}, {Pogge}, {Berg}, {Croxall}, {Bartlett}, {Arellano-C{\'o}rdova}, \& {Moustakas}}]{Rogers2022}
{Rogers}, N. S.~J., {Skillman}, E.~D., {Pogge}, R.~W., {et~al.} 2022, \apj, 939, 44, \dodoi{10.3847/1538-4357/ac947d}

\bibitem[{Rogers {et~al.}(2024)Rogers, Strom, Rudie, Trainor, Raptis, \& von Raesfeld}]{Rogers2024}
Rogers, N. S.~J., Strom, A.~L., Rudie, G.~C., {et~al.} 2024, The Astrophysical Journal Letters, 964, L12, \dodoi{10.3847/2041-8213/ad2f37}

\bibitem[{{Rogers} {et~al.}(2025){Rogers}, {Strom}, {Rudie}, {Trainor}, {von Raesfeld}, {Raptis}, {Korhonen Cuestas}, {Miller}, {Steidel}, {Maseda}, {Chen}, \& {Law}}]{Rogers2025}
{Rogers}, N. S.~J., {Strom}, A.~L., {Rudie}, G.~C., {et~al.} 2025, arXiv e-prints, arXiv:2509.18257, \dodoi{10.48550/arXiv.2509.18257}

\bibitem[{{Runco} {et~al.}(2021){Runco}, {Shapley}, {Sanders}, {Topping}, {Kriek}, {Reddy}, {Coil}, {Mobasher}, {Siana}, {Freeman}, {Shivaei}, {Azadi}, {Price}, {Leung}, {Fetherolf}, {de Groot}, {Zick}, {Fornasini}, \& {Barro}}]{Runco2021}
{Runco}, J.~N., {Shapley}, A.~E., {Sanders}, R.~L., {et~al.} 2021, \mnras, 502, 2600, \dodoi{10.1093/mnras/stab119}

\bibitem[{{Rynkun} {et~al.}(2019){Rynkun}, {Gaigalas}, \& {J{\"o}nsson}}]{Rynkun2019}
{Rynkun}, P., {Gaigalas}, G., \& {J{\"o}nsson}, P. 2019, \aap, 623, A155, \dodoi{10.1051/0004-6361/201834931}

\bibitem[{{Sanders} {et~al.}(2023){Sanders}, {Shapley}, {Topping}, {Reddy}, \& {Brammer}}]{Sanders2023}
{Sanders}, R.~L., {Shapley}, A.~E., {Topping}, M.~W., {Reddy}, N.~A., \& {Brammer}, G.~B. 2023, \apj, 955, 54, \dodoi{10.3847/1538-4357/acedad}

\bibitem[{{Sanders} {et~al.}(2024){Sanders}, {Shapley}, {Topping}, {Reddy}, \& {Brammer}}]{Sanders2024}
---. 2024, \apj, 962, 24, \dodoi{10.3847/1538-4357/ad15fc}

\bibitem[{Sanders {et~al.}(2015)Sanders, Shapley, Kriek, Reddy, Freeman, Coil, Siana, Mobasher, Shivaei, Price, \& de~Groot}]{Sanders2015}
Sanders, R.~L., Shapley, A.~E., Kriek, M., {et~al.} 2015, The Astrophysical Journal, 799, 138, \dodoi{10.1088/0004-637X/799/2/138}

\bibitem[{{Sanders} {et~al.}(2016){Sanders}, {Shapley}, {Kriek}, {Reddy}, {Freeman}, {Coil}, {Siana}, {Mobasher}, {Shivaei}, {Price}, \& {de Groot}}]{Sanders2016}
{Sanders}, R.~L., {Shapley}, A.~E., {Kriek}, M., {et~al.} 2016, \apj, 816, 23, \dodoi{10.3847/0004-637X/816/1/23}

\bibitem[{Sanders {et~al.}(2019)Sanders, Jones, Shapley, Reddy, Kriek, Coil, Siana, Mobasher, Shivaei, Price, Freeman, Azadi, Leung, Fetherolf, Zick, Groot, Barro, \& Fornasini}]{Sanders2020a}
Sanders, R.~L., Jones, T., Shapley, A.~E., {et~al.} 2019, The Astrophysical Journal Letters, 888, L11, \dodoi{10.3847/2041-8213/ab5d40}

\bibitem[{{Sanders} {et~al.}(2021){Sanders}, {Shapley}, {Jones}, {Reddy}, {Kriek}, {Siana}, {Coil}, {Mobasher}, {Shivaei}, {Dav{\'e}}, {Azadi}, {Price}, {Leung}, {Freeman}, {Fetherolf}, {de Groot}, {Zick}, \& {Barro}}]{Sanders2021}
{Sanders}, R.~L., {Shapley}, A.~E., {Jones}, T., {et~al.} 2021, \apj, 914, 19, \dodoi{10.3847/1538-4357/abf4c1}

\bibitem[{Sanders {et~al.}(2025)Sanders, Shapley, Topping, Reddy, Berg, Khostovan, Bouwens, Brammer, Carnall, Cullen, Davé, Dunlop, Ellis, Schreiber, Furlanetto, Glazebrook, Illingworth, Jones, Kriek, McLeod, McLure, Narayanan, Oesch, Pahl, Pettini, Schaerer, Stark, Steidel, Tang, Clarke, Donnan, \& Kehoe}]{Sanders2025}
Sanders, R.~L., Shapley, A.~E., Topping, M.~W., {et~al.} 2025, The AURORA Survey: High-Redshift Empirical Metallicity Calibrations from Electron Temperature Measurements at z=2-10.
\newblock \doarXiv{2508.10099}

\bibitem[{{Sanders} {et~al.}(2025){Sanders}, {Shapley}, {Topping}, {Reddy}, {Berg}, {Bouwens}, {Brammer}, {Carnall}, {Cullen}, {Dav{\'e}}, {Dunlop}, {Ellis}, {F{\"o}rster Schreiber}, {Furlanetto}, {Glazebrook}, {Illingworth}, {Jones}, {Kriek}, {McLeod}, {McLure}, {Narayanan}, {Oesch}, {Pahl}, {Pettini}, {Schaerer}, {Stark}, {Steidel}, {Tang}, {Clarke}, {Donnan}, \& {Kehoe}}]{Sanders2024b}
{Sanders}, R.~L., {Shapley}, A.~E., {Topping}, M.~W., {et~al.} 2025, \apj, 989, 209, \dodoi{10.3847/1538-4357/adf066}

\bibitem[{{Scholte} {et~al.}(2025){Scholte}, {Cullen}, {Carnall}, {Arellano-C{\'o}rdova}, {Stanton}, {Barrufet}, {Begley}, {Bondestam}, {Donnan}, {Dunlop}, {Leung}, {McLeod}, {McLure}, {Moustakas}, {Pollock}, {Shapley}, {Stevenson}, \& {Zou}}]{Scholte2025}
{Scholte}, D., {Cullen}, F., {Carnall}, A.~C., {et~al.} 2025, \mnras, 540, 1800, \dodoi{10.1093/mnras/staf834}

\bibitem[{Schreiber {et~al.}(2019)Schreiber, Übler, Davies, Genzel, Wisnioski, Belli, Shimizu, Lutz, Fossati, Herrera-Camus, Mendel, Tacconi, Wilman, Beifiori, Brammer, Burkert, Carollo, Davies, Eisenhauer, Fabricius, Lilly, Momcheva, Naab, Nelson, Price, Renzini, Saglia, Sternberg, Dokkum, \& Wuyts}]{Schreiber2019}
Schreiber, N. M.~F., Übler, H., Davies, R.~L., {et~al.} 2019, The Astrophysical Journal, 875, 21, \dodoi{10.3847/1538-4357/ab0ca2}

\bibitem[{{Searle} \& {Sargent}(1972)}]{Searle1972}
{Searle}, L., \& {Sargent}, W. L.~W. 1972, \apj, 173, 25, \dodoi{10.1086/151398}

\bibitem[{{Shapley} {et~al.}(2005){Shapley}, {Steidel}, {Erb}, {Reddy}, {Adelberger}, {Pettini}, {Barmby}, \& {Huang}}]{Shapley2005}
{Shapley}, A.~E., {Steidel}, C.~C., {Erb}, D.~K., {et~al.} 2005, \apj, 626, 698, \dodoi{10.1086/429990}

\bibitem[{{Shapley} {et~al.}(2015){Shapley}, {Reddy}, {Kriek}, {Freeman}, {Sanders}, {Siana}, {Coil}, {Mobasher}, {Shivaei}, {Price}, \& {de Groot}}]{Shapley2015}
{Shapley}, A.~E., {Reddy}, N.~A., {Kriek}, M., {et~al.} 2015, \apj, 801, 88, \dodoi{10.1088/0004-637X/801/2/88}

\bibitem[{Shapley {et~al.}(2019)Shapley, Sanders, Shao, Reddy, Kriek, Coil, Mobasher, Siana, Shivaei, Freeman, Azadi, Price, Leung, Fetherolf, Groot, Zick, Fornasini, \& Barro}]{Shapley2019}
Shapley, A.~E., Sanders, R.~L., Shao, P., {et~al.} 2019, The Astrophysical Journal Letters, 881, L35, \dodoi{10.3847/2041-8213/ab385a}

\bibitem[{{Shapley} {et~al.}(2025){Shapley}, {Sanders}, {Topping}, {Reddy}, {Berg}, {Bouwens}, {Brammer}, {Carnall}, {Cullen}, {Dav{\'e}}, {Dunlop}, {Ellis}, {F{\"o}rster Schreiber}, {Furlanetto}, {Glazebrook}, {Illingworth}, {Jones}, {Kriek}, {McLeod}, {McLure}, {Narayanan}, {Oesch}, {Pahl}, {Pettini}, {Schaerer}, {Stark}, {Steidel}, {Tang}, {Clarke}, {Donnan}, \& {Kehoe}}]{Shapley2025}
{Shapley}, A.~E., {Sanders}, R.~L., {Topping}, M.~W., {et~al.} 2025, \apj, 980, 242, \dodoi{10.3847/1538-4357/adad68}

\bibitem[{{Stanton} {et~al.}(2024){Stanton}, {Cullen}, {McLure}, {Shapley}, {Arellano-C{\'o}rdova}, {Begley}, {Amor{\'\i}n}, {Barrufet}, {Calabr{\`o}}, {Carnall}, {Cirasuolo}, {Dunlop}, {Donnan}, {Hamadouche}, {Liu}, {McLeod}, {Pentericci}, {Pozzetti}, {Sanders}, {Scholte}, \& {Topping}}]{Stanton2024}
{Stanton}, T.~M., {Cullen}, F., {McLure}, R.~J., {et~al.} 2024, \mnras, 532, 3102, \dodoi{10.1093/mnras/stae1705}

\bibitem[{{Stanton} {et~al.}(2025{\natexlab{a}}){Stanton}, {Cullen}, {Carnall}, {Scholte}, {Arellano-C{\'o}rdova}, {McLeod}, {Begley}, {Donnan}, {Dunlop}, {Hamadouche}, {McLure}, {Shapley}, {Bondestam}, \& {Stevenson}}]{Stanton2025}
{Stanton}, T.~M., {Cullen}, F., {Carnall}, A.~C., {et~al.} 2025{\natexlab{a}}, \mnras, 537, 1735, \dodoi{10.1093/mnras/staf106}

\bibitem[{{Stanton} {et~al.}(2025{\natexlab{b}}){Stanton}, {Cullen}, {Carnall}, {Scholte}, {Arellano-C{\'o}rdova}, {Shapley}, {McLeod}, {Donnan}, {Begley}, {Dav{\'e}}, {Dunlop}, {McLure}, {Rowlands}, {Bondestam}, {Hamadouche}, {Leung}, {Stevenson}, \& {Taylor}}]{Stanton2025b}
---. 2025{\natexlab{b}}, arXiv e-prints, arXiv:2511.00705, \dodoi{10.48550/arXiv.2511.00705}

\bibitem[{{Steidel} {et~al.}(2016){Steidel}, {Strom}, {Pettini}, {Rudie}, {Reddy}, \& {Trainor}}]{Steidel2016}
{Steidel}, C.~C., {Strom}, A.~L., {Pettini}, M., {et~al.} 2016, \apj, 826, 159, \dodoi{10.3847/0004-637X/826/2/159}

\bibitem[{{Steidel} {et~al.}(2014){Steidel}, {Rudie}, {Strom}, {Pettini}, {Reddy}, {Shapley}, {Trainor}, {Erb}, {Turner}, {Konidaris}, {Kulas}, {Mace}, {Matthews}, \& {McLean}}]{Steidel2014}
{Steidel}, C.~C., {Rudie}, G.~C., {Strom}, A.~L., {et~al.} 2014, \apj, 795, 165, \dodoi{10.1088/0004-637X/795/2/165}

\bibitem[{{Stiavelli} {et~al.}(2025){Stiavelli}, {Morishita}, {Chiaberge}, {Leethochawalit}, {Norman}, {Ricotti}, {Roberts-Borsani}, {Treu}, {Vanzella}, {Wyse}, {Zhang}, \& {Boyett}}]{Stiavelli2025}
{Stiavelli}, M., {Morishita}, T., {Chiaberge}, M., {et~al.} 2025, \apj, 981, 136, \dodoi{10.3847/1538-4357/adb5f3}

\bibitem[{{Storey} {et~al.}(2014){Storey}, {Sochi}, \& {Badnell}}]{Storey2014}
{Storey}, P.~J., {Sochi}, T., \& {Badnell}, N.~R. 2014, \mnras, 441, 3028, \dodoi{10.1093/mnras/stu777}

\bibitem[{Strom {et~al.}(2022)Strom, Rudie, Steidel, \& Trainor}]{Strom2022}
Strom, A.~L., Rudie, G.~C., Steidel, C.~C., \& Trainor, R.~F. 2022, The Astrophysical Journal, 925, 116, \dodoi{10.3847/1538-4357/ac38a3}

\bibitem[{{Strom} {et~al.}(2017){Strom}, {Steidel}, {Rudie}, {Trainor}, {Pettini}, \& {Reddy}}]{Strom2017}
{Strom}, A.~L., {Steidel}, C.~C., {Rudie}, G.~C., {et~al.} 2017, \apj, 836, 164, \dodoi{10.3847/1538-4357/836/2/164}

\bibitem[{Strom {et~al.}(2023)Strom, Rudie, Trainor, Brammer, Maseda, Raptis, Rogers, Steidel, Chen, \& Law}]{Strom2023}
Strom, A.~L., Rudie, G.~C., Trainor, R.~F., {et~al.} 2023, The Astrophysical Journal Letters, 958, L11, \dodoi{10.3847/2041-8213/ad07dc}

\bibitem[{{Tayal} \& {Zatsarinny}(2010)}]{Tayal2010}
{Tayal}, S.~S., \& {Zatsarinny}, O. 2010, \apjs, 188, 32, \dodoi{10.1088/0067-0049/188/1/32}

\bibitem[{Thomas {et~al.}(2005)Thomas, Maraston, Bender, \& de~Oliveira}]{Thomas2005}
Thomas, D., Maraston, C., Bender, R., \& de~Oliveira, C.~M. 2005, The Astrophysical Journal, 621, 673, \dodoi{10.1086/426932}

\bibitem[{{Tinsley}(1979)}]{Tinsley1979}
{Tinsley}, B.~M. 1979, \apj, 229, 1046, \dodoi{10.1086/157039}

\bibitem[{{Topping} {et~al.}(2020){Topping}, {Shapley}, {Reddy}, {Sanders}, {Coil}, {Kriek}, {Mobasher}, \& {Siana}}]{Topping2020}
{Topping}, M.~W., {Shapley}, A.~E., {Reddy}, N.~A., {et~al.} 2020, \mnras, 499, 1652, \dodoi{10.1093/mnras/staa2941}

\bibitem[{{Topping} {et~al.}(2024){Topping}, {Stark}, {Senchyna}, {Plat}, {Zitrin}, {Endsley}, {Charlot}, {Furtak}, {Maseda}, {Smit}, {Mainali}, {Chevallard}, {Molyneux}, \& {Rigby}}]{Topping2024}
{Topping}, M.~W., {Stark}, D.~P., {Senchyna}, P., {et~al.} 2024, \mnras, 529, 3301, \dodoi{10.1093/mnras/stae682}

\bibitem[{{Topping} {et~al.}(2025){Topping}, {Sanders}, {Shapley}, {Pahl}, {Reddy}, {Stark}, {Berg}, {Clarke}, {Cullen}, {Dunlop}, {Ellis}, {Schreiber}, {Illingworth}, {Jones}, {Narayanan}, {Pettini}, \& {Schaerer}}]{Topping2025}
{Topping}, M.~W., {Sanders}, R.~L., {Shapley}, A.~E., {et~al.} 2025, \mnras, 541, 1707, \dodoi{10.1093/mnras/staf903}

\bibitem[{{Velichko} {et~al.}(2024){Velichko}, {De Cia}, {Konstantopoulou}, {Ledoux}, {Krogager}, \& {Ramburuth-Hurt}}]{Velichko2024}
{Velichko}, A., {De Cia}, A., {Konstantopoulou}, C., {et~al.} 2024, \aap, 685, A103, \dodoi{10.1051/0004-6361/202348601}

\bibitem[{{Wallerstein}(1962)}]{Wallerstein1962}
{Wallerstein}, G. 1962, \apjs, 6, 407, \dodoi{10.1086/190067}

\bibitem[{{Wang} {et~al.}(2014){Wang}, {Ferland}, {Lykins}, {Porter}, {van Hoof}, \& {Williams}}]{Wang2014}
{Wang}, Y., {Ferland}, G.~J., {Lykins}, M.~L., {et~al.} 2014, \mnras, 440, 3100, \dodoi{10.1093/mnras/stu514}

\bibitem[{{Welch} {et~al.}(2025){Welch}, {Rivera-Thorsen}, {Rigby}, {Hutchison}, {Olivier}, {Berg}, {Sharon}, {Dahle}, {Owens}, {Bayliss}, {Khullar}, {Chisholm}, {Hayes}, \& {Kim}}]{Welch2025}
{Welch}, B., {Rivera-Thorsen}, T.~E., {Rigby}, J.~R., {et~al.} 2025, \apj, 980, 33, \dodoi{10.3847/1538-4357/ada76c}

\bibitem[{Wiese {et~al.}(1996)Wiese, Fuhr, \& Deters}]{Wiese1996}
Wiese, W., Fuhr, J., \& Deters, T. 1996, JPCRD, Monograph 7, Atomic transition probabilities of carbon, nitrogen, and oxygen: a critical data compilation,  NY: AIP Press

\bibitem[{{Zeippen}(1982)}]{Zeippen1982}
{Zeippen}, C.~J. 1982, \mnras, 198, 111, \dodoi{10.1093/mnras/198.1.111}

\end{thebibliography}
\clearpage

\end{document}